\documentclass[twocolumn,showpacs,preprintnumbers,amsmath,amssymb]{revtex4}

\usepackage{graphicx}
\usepackage{dcolumn}
\usepackage{bm}
\usepackage{color}
\usepackage{amsmath}
\usepackage{amssymb}
\usepackage{dsfont}
\usepackage{color}
\usepackage{calc}
\usepackage{hyperref}
\usepackage{multirow}
\usepackage{subfig}
\usepackage{float}
\usepackage[normalem]{ulem}
\usepackage{natmove}

\def\be{\begin{equation}}
\def\ee{\end{equation}}
\def\bea{\begin{eqnarray}}
\def\eea{\end{eqnarray}}

\graphicspath{{./figures/}}

\begin{document}

\title{Quantum-phase-field: from de Broglie - Bohm double solution program to doublon networks}

\author{  J.~Kundin}
 \email{julia.kundin@rub.de}
\author{  I.~Steinbach}
 \email{ingo.steinbach@rub.de}
\affiliation{ Ruhr-University Bochum, ICAMS, Universitaetsstrasse 150, 44801 Bochum, Germany}



\date{\today}

\begin{abstract}
We study different forms of linear and non-linear field equations, so-called `phase-field' equations, in relation to the de~Broglie-Bohm double solution program. This defines a framework in which elementary particles are described by localized non-linear wave solutions moving by the guidance of a pilot wave, defined by the solution of a Schr\"odinger type equation. First, we consider the phase-field order parameter as the phase for the linear pilot wave, second as the pilot wave itself and third as a moving soliton interpreted as a massive particle. In the last case, we introduce the equation for a superwave, the amplitude of which can be considered as a particle moving in accordance to the de~Broglie-Bohm theory. Lax pairs for the coupled problems are constructed in order to discover possible non-linear equations which can describe the moving particle and to propose a framework for investigating coupled solutions. Finally, doublons in 1+1 dimensions are constructed as self similar solutions of a non-linear phase-field equation forming a finite space-object. Vacuum quantum oscillations within the doublon determine the evolution of the coupled system.  Applying a conservation constraint and using general symmetry considerations, the doublons are arranged as a network in 1+1+2 dimensions where nodes are interpreted as elementary particles. A canonical procedure is proposed to treat charge and electromagnetic exchange.
\end{abstract}


\maketitle

\section{Introduction}\label{Intro}

Since the emergence of quantum theory physicists try to understand the physical world as a dual phenomenon of waves and particles. A very promising concept, but today almost forgotten,
 is the double solution program of de Broglie and Bohm (dBB) \cite{Bohm1952a,Bohm1952b,deBroglie1960,deBroglie1964,deBroglie1971}. The concept is based on two coupled wave equations. The so-called `pilot wave'  corresponds to the probability wave, $\psi$, in the Copenhagen interpretation of quantum mechanics and evolves according to a linear Schr\"odinger type equation. This pilot wave couples to the $u$ wave (in de Broglies notation \cite{deBroglie1960}), which shall describe physical `particles'. Doing this, it is possible to interpret e.g. the famous `double slit' experiment in a picture where the particle remains as such from the moment of emission until the moment of absorption, but is `informed' about the probability of different paths by the pilot wave, i.e. it is guided by the pilot wave \cite{Zeh1999}. The path of the particle within the period between emission and absorption is unknown (if no additional measurement is performed which would spoil the experiment), that is why the particle itself is termed `hidden' during the experiment. The theory is `non-local, realistic' along the classification of Bell \cite{Bell1964}.  Non-local means that physical observables are represented by gradient operators acting on fields in space and time. Realistic means that physical observables have well defined values (not necessarily `exact', but subject to Heisenberg's uncertainty) independent of a measurement. This contrasts to `local, realistic' classical theories, Newton's mechanics and General Relativity (GR), and the `non-local, probabilistic' Copenhagen interpretation of quantum mechanics.

 De~Broglie, the father of the wave interpretation of elementary particles, presented his ideas, how to construct stable elementary particles and their interaction as wave phenomena, first at the Solvay conference 1927. For a recent review and details see \cite{Willox2017}. The solution of a linear Schr\"odinger type wave equation couples to the solution of a non-linear wave equation which has localized amplitudes interpreted as massive particles,  the $u$ wave. Both waves are connected by their phases, which is why the pilot wave can guide the particle. According to Bohm's work, the Schr\"odinger equation for the $\psi$  wave can be decomposed into two equations,  for the amplitude ($R$ in Bohm's notation) and for the phase ($S$ in Bohm's notation). The probability of finding a particle ($P=R^2$) relates to the probability in the Copenhagen interpretation. Using the phase $S$, the velocity of the particle  can be found as $v_{\rm  p}= -c^2\nabla S/\partial_tS$. This can be seen in analogy to the particles in water waves which make loops that gradually advance in the direction of wave propagation \cite{Constantin2006}.

Why treat matter as coupled waves? Why not simply accept the existence of elementary particles as point masses? The most prominent opponent of the dualism of mass and space was Einstein, who worked for long on a `unitary'  or `monistic' view of the physical world, as reported in \cite{Willox2017}:

\textit{"[...] Einstein [...] had a very similar objective in the framework of his theory of general relativity in the 20's: the postulate of geodesics would not be an extra hypothesis, but would be obeyed, de facto, by peaked solutions of Einstein's nonlinear equations moving on a weakly varying metric background. This idea presents many deep similarities with de~Broglie's guidance equation which lies at the heart of the de~Broglie-Bohm hidden variable theory \cite{Bohm1952a,Bohm1952b}. In fact, Bohm-de~Broglie trajectories are the counterpart of geodesic trajectories in Einstein's unitarian version of general relativity ..."}

Today it is almost fully accepted that elementary particles cannot be considered as point masses and a non-local theory, as the dBB double solution program, is needed. An appropriate quantum field theory, which is based on elementary space quanta with finite dimensions, e.g. strings \cite{Zwiebach2009} or branes \cite{Zaslow2008},  had not been worked out at that time, and is still missing in a concise form today.  The double solution program offers an elegant solution to this problem because it combines a linear Schr\"odinger type wave equation for a $\psi$-wave with a non-linear wave equation for a $u$-wave describing the particle. 
 
The main difficulty, why the theory did not prevail, was the construction of the $u$ wave representing the particle, or other quantum-mechanical objects. Although de~Broglie's primary idea had been the description of matter as a wave phenomenon, the double solution program is based on the traditional understanding of `particles' and `space' as separate elements of matter. Recently, the second author developed a monistic view of matter, called `quantum-phase-field concept' \cite{Steinbach2017zn,footnote1}, treating space-time and mass as two manifestations of matter with dualistic character. In the quantum-phase-field concept, a special field solution, which we will call `doublon' in the following, forms the elementary building block of matter. It unifies the aspects of particles and space. Only after publication, his attention was called to the similarity with the dBB double solution program, that shall be elaborated in the present article. 


Here we have to link to another development, which began in the 1960s in the analysis of non-linear wave equations that are related to a completely different phenomenon: shallow water waves in channels and ocean shores. The problem has been described by Korteweg and de~Vries (KdV) in 1896 \cite{Korteweg1895}. Solutions of this type of non-linear wave equations are characterized by a self-similar shape of wave packets, nowadays known as `solitons', for review see \cite{Scott1973,Manton2008}. The soliton solutions had been unknown to de~Broglie and Bohm in the beginning of the 1960s.

Today the `soliton' appears in a new application, the phase-field (PF) theory  \cite{boettinger_phase-field_2002,Steinbach2009,Steinbach2013} which bases on a soliton solution of a special non-linear wave equation derived from a Ginzburg-Landau functional. The PF theory is applied to investigate pattern formation in condensed matter physics and materials science. The pattern here is the morphology of a `phase' in real space and its evolution, therefrom the name. A phase is identified by an order parameter $\phi$, called `phase-field' variable. It is a state of matter with a discontinuous property relation to the other phases in the sense of Landau \cite{Landau1959}. We shall not confuse this interpretation of a `phase' with the phase of a wave, although we will show that indeed it can have a similar formal meaning if we speak about localized wave solutions, i.e. solitons.

One distinguishes primarily two sets of 'solitons', as depicted in Figure \ref{soliton} with the eigen-coordinate $\xi$, the space coordinate $x$, time coordinate $t$, the size $\eta$ and the velocity $v$:
\begin{figure}[ht]
 \centering
\includegraphics[width=8.5cm]{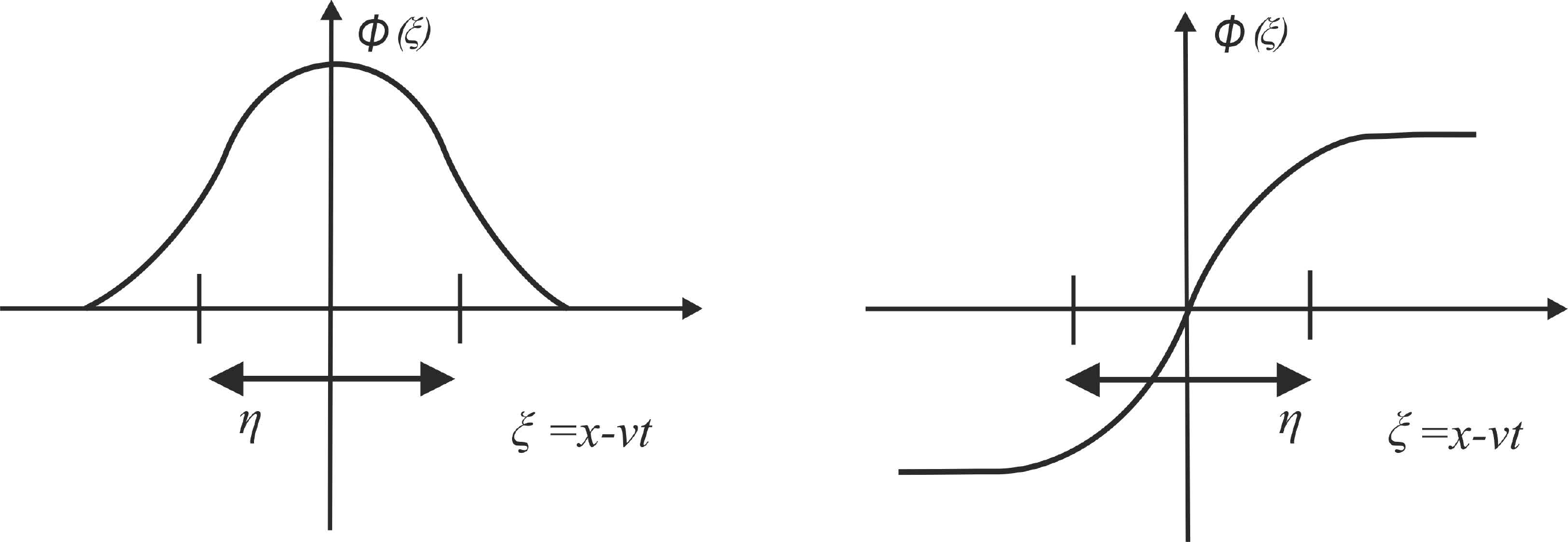}
\caption{The soliton in $1+1$ dimensions. a) Symmetric soliton, b) Half-sided soliton as an integral form of the symmetric soliton. $\eta$ is the characteristic size of the wave. }\label{soliton}
\end{figure}

Figure \ref{soliton} a) depicts the classical soliton as an excitation against a homogeneous background. Besides numerous application in wave mechanics (see \cite{Scott1973} for an exhaustive review at its time), it has been proposed as a template to describe elementary particles \cite{Manton2008}. The original soliton is a 1-dimensional, or planar 2-dimensional solution of a spacial non-linear wave equation. True 3-dimensional realizations have been shown to be mathematically impossible \cite{Craig2002}. The construction of `particles' then is only possible in a spherical symmetric approximation (see e.g. \cite{Bohun1999}), which by construction, prevents investigating scattering events. It has also been realized early on, that the intrinsic length scale of the soliton, its size $\eta$, leads to a `natural mode of quantization' \cite{Olsen1974}. The latter statement leads back to the de~Broglie-Bohm program. This connection has been little addressed in the literature, but is clearly pointed out by  Colin et al. \cite{Willox2017}.  

Figure \ref{soliton} b) shows a solution of an integral form of the classical Korteweg-de~Vries type equation, which introduces parity in the space coordinate. We will later, chapter \ref{section_doublon}, combine two of these `half-sided solitons', a 'right moving soliton' and a `left moving soliton', to an antisymmetric pair, which we call `doublon' \cite{footnote2} The doublon will serve i) to define massive particles with positive energy by gradient operators and ii) define `space' as the distance between the particles, attributed by negative energy.
\\

We will proceed as follows: 
\begin {itemize}
\item Describe the relation of the `phase-field` equation to  the dBB double solution program.
\item Present Lax-pairs which couple two wave functions similar to the dBB program.
\item Construct `doublons' from an antisymmetric pair of half-sided left /right moving solitons from the `phase-field` equation.
\item Construct a `doublon network' in 1+1+2 dimensions which embeds massive particles.
\item Propose a generalization of the concept to charged particles and fields.
\item Discuss limitations and perspectives of the current approach.
\end {itemize}

\section{PF as a toy problem for the dBB double solution program}

 \subsection{ PF equation for a phase of a wave function } \label{Section_DeBroglieEquations_Phase}
 
The phase-field order parameter $\phi$ can be considered in analogy to the phase $S$ in the dBB double solution program \cite{deBroglie1971}.  De~Broglie and Bohm derived equations for the phase and for the amplitude taking the pilot wave in the form $\psi=Re^{iS}$ and substituting it into the Schr\"odinger equation. Here we solve the inverse problem and  seek for an equation for a $\psi$ wave, whose phase evolves according to a PF equation. In order to be consistent, we define $\phi$ as the phase of  a  wave function $\psi$ with an amplitude $a_\psi$ as
   \begin{align}\label{v_wave}
 \psi=a_\psi e^{i \phi}.
 \end{align}

Then we consider the linear PF equation as an evolution equation for the phase:
 \begin{eqnarray}\label{pfe0}
\tau \partial_t\phi = \eta^2\partial_{xx} \phi + \phi+\eta e\partial_x\phi, 
\end{eqnarray}
where $\partial_t = \frac \partial {\partial t}$, $\partial_x = \frac \partial {\partial x}$ and $\partial_{xx} = \frac {\partial^2} {\partial x^2}$.

Here $\tau$ is a kinetic parameter with dimension of time,  $\eta$ is a characteristic length to be identified with the size of the resulting soliton solution (see Figure \ref{soliton}) which is also called `interface width' in the PF model, and $e$ is a dimensionless parameter related to the velocity of transport. The simple solution of this equation is $\phi=e^ {i\theta}$ with $\theta =\dfrac{x+v_\phi t}{\eta}$ and the velocity $v_\phi = \dfrac{\eta e}{\tau}$. For this kind of solution, the PF equation can be separated into two equations: the advection equation $ \partial_t\phi =  v_\phi\partial_x\phi$ and the equation for the self-similar shape of the soliton.

From the equation for the phase, we can reconstruct the equation for $\psi$. The following relations can be useful as parts of the sought equation:
  \begin{align}\label{terms}
 \partial_t \psi &= \partial_t (a_\psi e^{i\phi})= \frac{\partial_ta_\psi }{a_\psi }\psi + i (\partial_t \phi)\psi \nonumber\\
 &=\frac{\partial_t(a_\psi ^2)}{2a_\psi ^2}\psi + i (\partial_t \phi)\psi, \nonumber\\
\partial_{xx}\psi &= 2i\frac{\partial_x a_\psi }{a_\psi }(\partial_x\phi) \psi + i^2(\partial_x \phi)^2\psi + i(\partial_{xx} \phi)\psi  +\frac{\partial_{xx} a_\psi }{a_\psi } \psi \nonumber\\
&=\left[-(\partial_x \phi)^2+\frac{\partial_{xx} a_\psi }{a_\psi }\right]\psi+ i \left[2\frac{\partial_{x} a_\psi }{a_\psi }(\partial_x \phi)+\partial_{xx} \phi\right]\psi.
\end{align}
%

With these relations, the following  form of the evolution equation for $\psi$ can be  suggested:
 \begin{eqnarray}\label{v_equation}
\tau \partial_t \psi = \eta^2\partial_{xx} \psi + i\phi \psi+i\eta e_0(\partial_x\phi )\psi,
\end{eqnarray}
where $e_0$ is a parameter to be defined. By the substitution of the solution \eqref{v_wave} in this equation and collecting the necessary imaginary terms, we obtain the equation for the phase \eqref{pfe0} with $e=e_0$.

The equation for the amplitude is the collection of the remaining terms, i.e.,
 \begin{eqnarray}\label{a_equation}
\tau \frac{\partial_t(a_\psi ^2)}{2a_\psi ^2} = -\eta^2(\partial_{x} \phi)^2 + \eta^2\frac{\partial_{xx} a_\psi }{a_\psi } +  2i\eta^2\frac{\partial_x a_\psi}{a_\psi}(\partial_x \phi).
\end{eqnarray}

After substitution $\dfrac{\partial_{xx}a_\psi }{a_\psi } = \dfrac{\partial_{xx}(a_\psi ^2)}{2a_\psi ^2}-\dfrac{(\partial_{x} a_\psi )^2}{a_\psi ^2}$, we obtain
 \begin{align}\label{a_equation2}
\tau\partial_t(a_\psi ^2) &=\eta^2\partial_{xx}(a_\psi ^2)-2 \eta^2(\partial_x \phi)^2a_\psi ^2  -2\eta^2(\partial_{x} a_\psi )^2\nonumber\\
&+  4i \eta^2 a_\psi(\partial_x a_\psi)(\partial_x \phi) . 
\end{align}

Since this equation can be considered as a PF-like equation, it can be divided into two equations
 \begin{align}\label{a_equation3}
\tau\partial_t(a_\psi ^2)&= 2i\eta^2(\partial_x \phi)\partial_x( a_\psi^2),\\
\eta^2\partial_{xx}(a_\psi ^2)&=2 \eta^2(\partial_x \phi)^2 a_\psi ^2 +2\eta^2\left(\frac{\partial_{x} a_\psi }{ a_\psi}\right)^2 a_\psi^2.\label{a_equation4}
\end{align}

Equation \eqref{a_equation3} is the advection equation 
 \begin{eqnarray}\label{a_equation5}
\partial_t (a_\psi ^2)=  v_{a_\psi}\partial_{x} ( a_\psi ^2)
\end{eqnarray}
with the velocity 
 \begin{eqnarray}\label{vel_R}
v_{a_\psi }=\dfrac{2i\eta^2}{\tau }\partial_{x}\phi.
\end{eqnarray}

Equation \eqref{a_equation4} defines the self-similar shape and can be written as
 \begin{eqnarray}
\eta_{a_\psi }^2\partial_{xx}(a_\psi ^2)= -a_\psi ^2\label{a_equation6}
\end{eqnarray}
with 
 \begin{eqnarray}\label{eta_a_psi}
 \eta_{a_\psi }^2=-\frac{1}{2}\left((\partial_x \phi)^2+\left(\dfrac{\partial_{x} a_\psi }{ a_\psi}\right)^2\right)^{-1}.
 \end{eqnarray}

 A solution of this equation is $a_\psi ^2(x,0)=e^{ix/\eta_{a_\psi }}$ or $a_\psi (x,0)=e^{ix/(2\eta_{a_\psi })}$.
 
Using the advection equation (\ref{a_equation3}), we obtain a total solution for the amplitude $ a_\psi ^2(x,t)=e^{i\theta_{a}}$ with $\theta_{a}~=~\dfrac{x+v_{a_\psi} t}{\eta_{a_\psi }}$.

Assuming that the velocity of the amplitude $v_{a_\psi}~=~\dfrac{2i\eta^2}{\tau }\partial_{x}\phi$ is equal to the particle velocity $v_{\rm p}~=~-c^2\dfrac{\partial_x \phi}{\partial_t \phi }$, and substituting $\partial_t \phi=\dfrac{ie\phi }{\tau}$, we can define the characteristic length $\eta$ from $\eta^2 = \dfrac{c^2\tau^2}{2e \phi}$. For the mean value $\bar \phi=\dfrac{1}{2}$, $\eta \approx \dfrac{c\tau}{\sqrt{e}}$. The  particle velocity is then defined as $v_{a_\psi }=\dfrac{2\phi c}{ \sqrt{e} }\approx\dfrac{c}{ \sqrt{e} }=v_{\rm p}$, and the velocity of the phase $\phi$ is defined as $v_\phi = c\sqrt{e}$. Moreover, by substitution in \eqref{eta_a_psi} $\dfrac{\partial_{x} a_\psi  }{ a_\psi }=\dfrac{i}{2\eta_{a_\psi }}$ and $\partial_x \phi =\dfrac{i\bar\phi}{\eta}$, we get $\eta_{a_\psi }= \eta$.  

According   to our results, the particle velocity decreases with increasing $e$ whereas the phase velocity increases. Both velocities  approach $c$  when $e$ decreases, moreover, $v_{\rm {a_\psi}}v_\phi=c^2$ in accordance with the dBB theory.

In quantum mechanics, the momentum (i.e., the particle velocity) should be inverse proportional to the  wave length, identified with the characteristic length $p\sim 1/\eta$, and direct proportional to the particle velosity. Hence, if we assume $\tau=\tau_0 e$, $\eta$ will be proportional to $\sqrt{e}$ and $p$ as well as $v_{\rm {a_\psi}}$ will be proportional to $1/\sqrt{e}$ according to expectations.  Based on that result, the characteristic length can be defined  as $\eta = c\tau_0\sqrt{e}$.

In summary, the particle is described by the amplitude function $a_\psi (x,t)$ as a one-dimensional singularity of size $\eta$  moving with the velocity $v_{a_\psi}$,  which is the solution of the PF-like equation \eqref{a_equation2}. This amplitude corresponds to the $u$-wave in the dBB theory and $\phi$ is the pilot wave for the amplitude.  The dynamics of $\phi$ on the other hand is governed by eq.~\eqref{pfe0}.

Finally, we show that eq.~\eqref{pfe0} can be transformed to the Schr\"odinger equation with a constant potential.  
Dividing the two first terms on the right hand side of eq.~\eqref{pfe0} by $2i$ (this is allowed because the PF equation can be separated into two separated equations), we get 
 \begin{align}\label{Schr_equation00}
 \tau\partial_t\phi &= \frac{\eta^2}{2i}\partial_{xx} \phi + \frac{\phi}{2i} +\eta e\partial_x\phi.
\end{align}

Then by substitution $ \tau_0=\dfrac{\hbar}{mc^2}$ and $\partial_x\phi=\dfrac{i\phi}{\eta}$, we obtain the Schr\"odinger-type equation
 \begin{align}\label{Schr_equation0}
 \partial_t\phi &= \frac{\hbar}{2im}\partial_{xx} \phi + \frac{iU_0}{\hbar}\phi
\end{align}
with $U_0=mc^2\left(1-\dfrac{1}{2e}\right)$. Here, $m$ is the mass of the particle, $\hbar$ it the Plank's constant.

 \subsection{PF equation for the pilot wave} \label{Section_DeBroglieEquations_Pilot}
 
 In this section, we consider the transformed PF equation \eqref{Schr_equation0} as  a Schr\"odinger-type equation with a given potential. The phase-field order parameter $\phi$ will be now considered as the pilot $\psi$ wave. We rewrite eq.~\eqref{Schr_equation00} as
 \begin{eqnarray}\label{Schr_equation1}
\tau \partial_t\psi = \frac{\eta^2}{2i}\partial_{xx} \psi + \frac{\psi}{2i}+\eta e\partial_x\psi
\end{eqnarray}
and define  $\psi=a_\psi  e^{iS}$ with $S =\dfrac{x+v_\psi t}{\eta}$ and $v_\psi~=~\dfrac{\partial_t S}{\partial_x S}~=~\dfrac{e\eta}{\tau}$. Then  we assume that the pilot wave $\psi$ is the phase of a superwave 
\begin{align}\label{v-wave}
 \Phi=a_\Phi e^{i\psi}.
\end{align}

 From eq.~\eqref{Schr_equation1}, we can reconstruct the equation for $\Phi $. Using the terms \eqref{terms} (where we replace $\psi$ by $\Phi $, $a_\psi$ by $a_\Phi $, and  $\phi$ by $\psi$), i.e.,
  \begin{align}\label{terms2}
 \partial_t \Phi  &= \frac{\partial_ta_\Phi }{a_\Phi }\Phi + i (\partial_t \psi)\Phi , \nonumber\\
\partial_{xx}\Phi  &=\left[-(\partial_x \psi)^2+\frac{\partial_{xx} a_\Phi }{a_\Phi }\right]\Phi \nonumber\\
&+ i \left[2\frac{\partial_{x} a_\Phi }{a_\Phi }(\partial_x \psi)+\partial_{xx} \psi\right]\Phi ,
\end{align}
 the following  form of the evolution equation for $\Phi $ can be  suggested:
 \begin{align}\label{v_equation2}
\tau \partial_t \Phi  &= \frac{\eta^2\partial_{xx} \Phi }{2i} + \frac{\psi \Phi}{2}  + \frac{ f(a_\Phi )\Phi }{2i} \nonumber\\
&+ i\eta e_0(\partial_x\psi )\Phi +\eta e_0^a\frac{ \partial_x a_\Phi } {a_\Phi }\Phi ,
\end{align}
where $e_0$ and $e_0^a$ are parameters responsible for the motion of $\psi$ and $a_\Phi$, respectively, which have to be defined. The term $\frac{ f(a_\Phi )\Phi }{i}$ is added by analogy to $\psi \Phi$. By substituting the $\Phi $ wave \eqref{v-wave} in this equation and collecting the imaginary terms  in \eqref{terms2}, the equation for $\psi$ \eqref{Schr_equation1} can be reconstructed with 
 \begin{eqnarray}\label{e_def2}
e=e_0-i\eta\frac{\partial_{x} a_\Phi }{a_\Phi }.
\end{eqnarray}

The equation for the amplitude is the collection of the real terms in \eqref{terms2}, i.e.,
 \begin{eqnarray}
\tau \frac{\partial_ta_\Phi }{a_\Phi } =  \frac{\eta^2}{2i}\frac{\partial_{xx} a_\Phi }{a_\Phi} +  \frac{ f(a_\Phi )}{2i} - \frac{\eta^2}{2i}(\partial_{x} \psi)^2+\eta e_0^a\frac{ \partial_xa_\Phi } {a_\Phi }. \nonumber\\\label{a_equation22}
\end{eqnarray}

This equation can be treated as a set of two equations
 \begin{align}
\tau\partial_ta_\Phi &=  \eta e_0^a \partial_xa_\Phi +  \frac{i \eta^2}{2}(\partial_{x} \psi)^2\left(\dfrac{a_\Phi }{\partial_{x} a_\Phi }\right){\partial_{x} a_\Phi } ,\label{a_equation32}\\
\eta^2_{a_\Phi}\partial_{xx}a_\Phi &=-a_\Phi f(a_\Phi ),\label{a_equation42}
\end{align}
where $\eta_{a_\Phi}=\eta$.
By assumption $f(a_\Phi ) =1$, a solution of eq.~\eqref{a_equation42} reads $a_\Phi (x,0)=e^{ix/\eta_{a_\Phi}}$.
Substituting $\dfrac{\partial_{x} a_\Phi }{a_\Phi }~=~\dfrac{i}{\eta}$, we can rewrite the advection equation \eqref{a_equation32} in the form
 \begin{align}\label{a_equation33}
\partial_ta_\Phi &=  v_{a_\Phi}\partial_{x} a_\Phi ,
\end{align}
with $v_{a_\Phi}=-\dfrac{\eta e^a}{\tau}$  and
 \begin{eqnarray}\label{e_a_def}
e^a=e_0^a+ \frac{\eta^2}{2}(\partial_{x} \psi)^2.
\end{eqnarray}

The full solution for the amplitude is $a_\Phi (x,t)=e^{i\theta_{a}} $ with
$\theta_{a}=\dfrac{x-v_{a_\Phi}t}{\eta_{a_\Phi}} $. The amplitude can be interpreted as a particle moving with the velocity $v_{a_\Phi}=v_{\rm p}=-c^2\dfrac{\partial_x S}{\partial_t S} $. From this, we can define the characteristic length as $\eta^2~=~\dfrac{c^2\tau^2}{ee^a}$. The particle velocity is then defined as $v_{a_\Phi}~=~c~\sqrt{\dfrac{e^a}{e } }$, and the velocity of the wave $\psi$ is defined as $v_\psi~=~c\sqrt{\dfrac{e}{e^a } }$. Here we have restriction for the particle velocity $e^a\leq e$.  By substituting  $\dfrac{\partial_{x} a_\Phi }{ a_\Phi}=\dfrac{i}{\eta_{a_\Phi}} $ in \eqref{e_def2}, we obtain $e=e_0+\eta/\eta_{a_\Phi}=e_0+1$.

Note that the phase and the amplitude of the wave $\psi =a_\psi e^{iS}$ evolve according to dBB equations \cite{deBroglie1971}, which are the Hamilton-Jacobi equation for the phase and the continuous equation for the amplitude. The difference between the dBB program and our solution is that in the present variant the amplitude of the superwave $a_\Phi $ is representing the particle which was previously considered as the $u$-wave. The guidance occurs as before by the dBB quantum force, which acts on the phase $S$, which then changes the particle velocity. Hence the function $\psi$ is the pilot wave for $a_\Phi $.

Finally, we can rewrite  eq.~\eqref{Schr_equation1} in physical units.  By substitution $\tau = \tau_0 e$, $\tau_0=\dfrac{\hbar}{m c^2}$, and $\eta=\tau_0v_\psi $, we obtain
 \begin{align}\label{Schr_equation_physU}
 \partial_t\psi &= \frac{\hbar v_\psi^2}{m c^2}\partial_{xx} \psi + \frac{mc^2}{\hbar }\psi+ v_\psi\partial_{x}\psi.
\end{align}

Then, multiplying the two first terms on the right hand side by $\dfrac{c^2}{(2i v_\psi^2)}$ and using $\partial_{x}\psi=i\dfrac{\psi}{\eta}$, we get the Schr\"odinger-type equation
 \begin{align}\label{Schr_equation_physU2}
 \partial_t\psi &= \frac{\hbar}{2im}\partial_{xx} \psi +  i\frac{U_0}{\hbar}\psi
\end{align}
with   $U_0=mc^2\left(1-\dfrac{c^2}{2v_\psi^2}\right)$. In the limit $v_\psi\rightarrow \infty$, $U_0\rightarrow mc^2$.

\section{Lax method for the solution of wave equations} \label{SectionLax}

The solution of a scattering problem in quantum mechanics is associated with the linear Lax operators $\mathcal{L}$ and $\mathcal{A}$, called Lax pair, which satisfy the Lax equation \cite{Lax1968},
\begin{align}\label{LaxEq0}
\mathcal{L}_t+[\mathcal{L},\mathcal{A}]=0.
\end{align}

This equation should reproduce a partial differential equation (PDE) for  a function $\phi$, whereas the Lax operators are applied to a function $\psi$, which is the solution of the Schr\"odinger-type equations. Therefore, the function $\psi$ is coupled  to the function $\phi$ by the Lax pair. Hence the Lax method can describe the coupling of two wave functions in the dBB double solution theory \cite{deBroglie1971}.

In order to find the solution of the PDE, one should solve the forward and inverse  scattering problems. The forward problem is solved by the first and second Lax equations with the eigenvalue $\lambda$:
\begin{align}
\mathcal{L}\psi=\lambda\psi,\label{Lax1}\\
\partial_t\psi=\mathcal{A}\psi.\label{Lax2}
\end{align}
The inverse scattering problem is solved by calculating the Gelfand-Levitan-Marchenko integral \eqref{GLM}. The function $\phi(x,t)$ is constructed from the formula
 \begin{align} \label{Result}
 \phi(x,t) = -2 \dfrac{\partial}{\partial x} K(x,x,t),
 \end{align}
 where $K(x,y,t)$ is the solution of the linear integral  equation
\begin{align}\label{GLM}
K(x,y)+F(x+y)+\int\limits_x^{+\infty} K(x,z)F(y+z)dz=0
\end{align}
defined for a fixed time $t$ and $F$ is the function which is based on the solution of the Lax equations for $\psi$.

\subsection{Lax pair for the phase and amplitude equations}
  
Now we show that the phase equation \eqref{pfe0}  can be reconstructed by the Lax method applied to a function $\psi$ whose evolution equation is described by the second Lax equation.

We choose the Lax operators as
\begin{align}\label{ZeroVariant}
\mathcal{L} &= \tau \eta\partial_x+ \tau \phi+\tau \eta i\partial_x+ \tau i\phi,\nonumber\\
\mathcal{A} &= \dfrac{\eta^2}{\tau}\partial_{xx} +i\dfrac{\phi}{\tau}+i\dfrac{\eta e_0}{\tau}\partial_x\phi.
\end{align}

The non-zero components of the Lax equation  \eqref{LaxEq0} have the terms
\begin{align}
&[ \phi,\partial_{xx}]\psi= \phi \psi_{xx}-\partial_{xx}(\phi\psi)\nonumber\\
&=\phi \psi_{xx}-\partial_x(\phi_x\psi+\phi\psi_x)\nonumber\\
&=\phi \psi_{xx}-(\phi_{xx}\psi+\phi_x\psi_x + \phi_x\psi_x+\phi\psi_{xx})\nonumber\\
&=-\phi_{xx}\psi-2\phi_x\psi_x,\nonumber\\
&[\partial_{x}, \phi]\psi=\partial_{x}(\phi\psi)-\phi\psi_x =\phi_x\psi,\nonumber\\
&[\partial_{x}, \phi_x]\psi=\phi_{xx}\psi.
\end{align}

Using $\phi=\phi_0 e^{i\theta}$ and  $\psi~=~\psi_0 e^{i\theta}$, we can substitute $\phi_x~=~(\log\phi)_x\phi$ and $\psi_x=(\log\psi)_x\psi$ and get $\phi_{xx}\psi~=~i\theta_x\phi_{x}\psi$, $\phi_{x}\psi=i\theta_x\phi\psi$, and $\phi_{x}\psi_x=i\theta_x\phi_{x}\psi$. Then we assume $\theta_x=1/\eta$. The components of the Lax equation including real and imaginary terms become
\begin{align}
&\eta^2[ \phi,\partial_{xx}]\psi=-\eta^2\phi_{xx}\psi-2\eta^2\phi_x\psi_x\nonumber\\
&=-\eta^2\phi_{xx}\psi-2i\eta\phi_x\psi,\nonumber\\
&\eta^2[ i\phi,\partial_{xx}]\psi=-i\eta^2\phi_{xx}\psi-2i\eta^2\phi_x\psi_x\nonumber\\
&=-i\eta^2\phi_{xx}\psi+2\eta\phi_x\psi,\nonumber\\
&\eta[\partial_{x}, i\phi]\psi=i\eta\phi_x\psi=-\phi\psi,\nonumber\\
&\eta[i\partial_{x}, i\phi]\psi=-\eta\phi_x\psi=-i\phi\psi,\nonumber\\
&\eta^2 e_0[\partial_{x}, i\phi_x]\psi= i\eta^2e_0\phi_{xx}\psi=-\eta e_0\phi_{x}\psi,\nonumber\\
&\eta^2 e_0[i\partial_{x}, i\phi_x]\psi= -\eta^2e_0\phi_{xx}\psi=-i\eta e_0\phi_{x}\psi.
\end{align}

After substitution in eq.~\eqref{LaxEq0}, we obtain
\begin{align}
 \tau\partial_{t}\phi &= \eta^2\partial_{xx}\phi+\phi+(e_0-2)\eta\partial_{x}\phi,\nonumber\\
  i\tau\partial_{t}\phi &= i\eta^2\partial_{xx}\phi+i\phi+i(e_0+2)\eta\partial_{x}\phi.
\end{align}

These equations recover the PF equation \eqref{pfe0} for $e~=~e_0\pm2$.

The first Lax equation for the $\psi$ functionwith the eigenvalues $\lambda_1$ and $\lambda_2$ is
\begin{align} \label{Psi_equations1}
 &i\eta\partial_{x}\psi + \phi\psi + \eta\partial_{x}\psi + i\phi\psi = \lambda_1 \psi +i\lambda_2 \psi,
\end{align}
  and second Lax equation is
\begin{align} \label{Psi_equations2} 
  \tau\partial_{t}\psi = \eta^2\partial_{xx}\psi +i\phi\psi+i\eta e_0(\partial_{x}\phi)\psi,
\end{align}
which  recovers eq.  \eqref{v_equation}  for the $\psi$ wave one to one.

The Lax pair can be found also for the waves in section \ref{Section_DeBroglieEquations_Phase} in the form $\phi= e^{i\theta}$ and  $\psi =a_{\psi} e^{i\phi}$, where $\phi$ is the phase of the pilot wave $\psi$. Doing so, we deviate from the Lax method, which requires that the phase of both functions should be equal and time independent. However, we use again the assumption that $\theta_x=1/\eta$. We show now that the following Lax pair can  reproduce  equations \eqref{pfe0} and  \eqref{a_equation2}:
\begin{align}\label{DoubleZeroVariant}
\mathcal{L} &= \tau \eta\partial_x+ \tau \phi+\tau \dfrac{2a_{\psi}(x,t)}{a_{\psi}}+\tau \eta i\partial_x+ \tau i \phi+\tau i\dfrac{2a_{\psi}(x,t)}{a_{\psi}},\nonumber\\
\mathcal{A} &= \dfrac{\eta^2}{\tau}\partial_{xx} +i\dfrac{\phi}{\tau}+i\dfrac{\eta e_0}{\tau}\partial_x\phi.
\end{align}

The non-zero components of the Lax equation have the terms
\begin{align}
&\eta^2[ \phi,\partial_{xx}]\psi=-\eta^2\phi_{xx}\psi-2\eta^2\phi_x\psi_x\nonumber\\
&=-\eta^2\phi_{xx}\psi-2i\eta^2(\phi_x)^2\psi-2\eta^2\frac{ \partial_x a_{\psi}}{a_{\psi}}\phi_x\psi,\nonumber\\
&\eta[\partial_{x}, i\phi]\psi=i\eta\phi_x\psi=-\phi\psi,\nonumber\\
&\eta^2 e_0[\partial_{x}, i\phi_x]\psi= i\eta^2e_0\phi_{xx}\psi=-\eta e_0\phi_{x}\psi,\nonumber\\
&\frac{\eta^2}{a_{\psi}}[ ia_{\psi}(x),\partial_{xx}]\psi =-i\eta^2\frac{\partial_{xx}a_{\psi}}{a_{\psi}}\psi-2i\eta^2\frac{\partial_x a_{\psi}}{a_{\psi}}\psi_x\nonumber\\
&=-i\eta^2\frac{\partial_{xx}a_{\psi}}{a_{\psi}}\psi-2i\eta^2 \frac{(\partial_xa_{\psi})^2}{a_{\psi}^2}\psi+2\eta^2\frac{\partial_xa_{\psi}}{a_{\psi}}\phi_x\psi.
\end{align}

 Here we used $ \psi_x = \dfrac{ \partial_x a_{\psi}}{a_{\psi}}\psi+i\phi_x \psi$ and normalized the $a_{\psi}$-operator by $a_{\psi}$. We also do not write the similar complex conjugate terms to be compact.

Substituting  all non-zero components in eq.~\eqref{LaxEq0} and collecting imaginary and real terms, we obtain two equations
\begin{align}
 \tau\partial_{t}\phi &= \eta^2\partial_{xx}\phi+\phi+e_0\eta(\partial_{x}\phi),\nonumber\\
  \tau\partial_{t}a_{\psi}^2 &= \eta^2\partial_{xx}a_{\psi}^2-2\eta^2(\partial_x \phi)^2a_{\psi}^2-2\eta^2(\partial_x a_{\psi})^2\nonumber\\
  &+4i\eta^2 a_{\psi}^2\frac{\partial_{x} a_{\psi}}{a_{\psi}}(\partial_{x}\phi),
\end{align}
which for $e=e_0$ lead to equations \eqref{pfe0} and  \eqref{a_equation2}  for the phase and the amplitude, whereas the evolution equation for the wave $\psi$ is in the same form as before (see eq.~\eqref{Psi_equations2}) i.e., recovers the equation  \eqref{v_equation} for the $\psi$ wave.  Hence, we have shown that the Lax method can be used also for the coupling of three functions, one of which is the function of two others. Note that in this case, $\phi$ is the pilot wave for $R$.

The Lax pair for eqs. \eqref{Schr_equation1}-\eqref{a_equation22} can be found in the  similar manner by adding the necessary terms in  the operators $\mathcal{L}$ and $\mathcal{A}$. Thus for the ansatz $\psi=e^{iS}$, $\Phi=a_\Phi e^{i\psi}$, and $\partial_x S =1/\eta$, the Lax pair for eqs. \eqref{Schr_equation1}-\eqref{a_equation22}  have the form:
\begin{align}\label{Variant105}
\mathcal{L} &= \tau \eta\partial_x+ \tau \psi+\tau \dfrac{2a_\Phi(x,t)}{a_\Phi}+\tau \eta i\partial_x+ \tau i \psi+\tau i \dfrac{2a_\Phi(x,t)}{a_\Phi},\nonumber\\
\mathcal{A} &= \dfrac{\eta^2}{2i\tau}\partial_{xx} +\dfrac{i\psi}{2i\tau}+i\dfrac{\eta e_0}{\tau}\partial_x\psi
+\dfrac{f(a_\Phi)}{2i\tau}+\dfrac{\eta e_0^a}{\tau }\dfrac{\partial_x a_\Phi}{ a_\Phi}.
\end{align}
Finally, the Lax equation \eqref{LaxEq0} recover eqs. \eqref{Schr_equation1} and \eqref{a_equation22} for $\psi$ and $a_\Phi$, and first and second Lax equations \eqref{Lax1}-\eqref{Lax2} recover eq.~\eqref{v_equation2}  for $\Phi$.
Note that in this case, the function $\psi$ is the pilot wave for $a_\Phi$.

\subsection{Lax pairs for the non-linear PF equation} \label{SubsectionLaxPF}
We consider a non-linear wave equation for a wave $\phi$, also known in material physics as the `phase-field equation' with double obstacle (DO) potential  \cite{Steinbach2009}
\begin{eqnarray}\label{pfe}
\tau \partial_{t}\phi = \eta^2\partial_{xx}\phi - f_\phi(\phi)+\eta e\partial_{x}\phi. 
\end{eqnarray}
$f_\phi(\phi)$ is a non-linear scalar function in $\phi$  which is the first derivative of the double obstacle potential
\begin{eqnarray}\label{dop}
f(\phi)=\frac{1}{2}|\phi(1-\phi)|.
\end{eqnarray}

The non-linearity is hidden in the break points $\phi~=~0$ and  $\phi = 1$. We will argue later that this ansatz is responsible for the localization of the wave solution. One can use here, without loss of generality, the functional form 
\begin{eqnarray}\label{dop2}
f_\phi(\phi)= \left(\frac{1}{2}-\phi\right){\rm signum}(\phi(1-\phi)),
\end{eqnarray}
that means $f_\phi(\phi)~=~(\frac{1}{2}-\phi)$ for $0\le \phi \le 1$ and $f_\phi(\phi)~=~(\phi-\frac{1}{2})$ otherwise.

A solution of  eq.~\eqref{pfe} with the potential \eqref{dop}  is, e.g.,
 \begin{align} \label{Sin_solution}
 \phi(x,t) = \frac{1}{2}\sin\left(\frac{ x-vt}{\eta}\right) +  \frac{1}{2},
\end{align}
for $- \frac \eta 2 -vt < x < \frac \eta 2 -vt$ for a half-wave traveling with velocity $v= \dfrac{\eta e}{\tau}$ (see also \cite{Steinbach2017zn}). We see that the non-linear absolute operator, $| . |$, cuts out one localized half wave from the periodic sinus wave. Although the solution is known, we will investigate the possibility to find a solution of this PDE by the `inverse scattering method' and demonstrate connections between two wave functions $\phi$ and $\psi$.
For this aim, we will find the Lax pairs and show that  eq.~\eqref{pfe} can be reproduced by the Lax representation.

\subsubsection{Variant I}

We choose the Lax operators in the form
\begin{align}\label{FirstVariant}
\mathcal{L} &= \tau \eta\partial_x+ \tau \phi,\nonumber\\
\mathcal{A} &= \dfrac{\eta^2}{\tau}\partial_{xx} +\dfrac{V(x)}{\tau\eta}+\dfrac{e_0}{\tau}\phi,
\end{align}
where $V(x) = \int^{x} {f_\phi(\phi)dx}$, $e_0$ is a constant to be defined.
This type of equations was suggested first  by Zakharov and  Shabat \cite{Zakharov1972}.

The non-zero components of the Lax equation are
\begin{align}
&[\partial_x, V(x)]\psi= \partial_x(V(x)\psi) -V(x)\psi_x\nonumber\\
&= \psi \partial_xV(x) +V(x)\psi_x-V(x)\psi_x=f_\phi(\phi)\psi;\nonumber\\
&[ \phi,\partial_{xx}]\psi=-\phi_{xx}\psi-2\phi_x\psi_x;\nonumber\\
&[\partial_{x}, \phi]\psi=\partial_{x}(\phi\psi)-\phi\psi_x =\phi_x\psi.
\end{align}

After substitution  in eq.~\eqref{LaxEq0}, we  obtain 
\begin{align}\label{PFeqLax}
\tau \partial_{t}\phi\psi + f_\phi(\phi)\psi- \eta^2\partial_{xx}\phi\psi-2\eta^2\partial_{x}\phi\partial_{x}\psi+e_0\eta\partial_{x}\phi\psi=0.
\end{align}

Then  we use $\psi_x=(\log\psi)_x\psi$, define a wave number $k=(\log\psi)_x$,   and obtain the phase-field equation in the form
\begin{align}
\tau \partial_{t}\phi = \eta^2\partial_{xx}\phi - f_\phi(\phi)+(2\eta k-e_0)\eta\partial_{x}\phi,
\end{align}
which recovers the phase-field equation \eqref{pfe} with $ e=(2\eta k-e_0)$.

Now we solve the forward scattering problem. The first Lax equation \eqref{Lax1}  after
substituting the operators and assuming $\lambda=\tau\eta k$ becomes
\begin{align} \label{LaxEq1}\partial_x\psi + \phi\psi/\eta=k\psi.
\end{align}

Assuming the asymptotic behavior
$\phi\rightarrow 0$ at $x\rightarrow - \infty$, we get
\begin{align}
\partial_x\psi=k\psi.
\end{align}

The solution of this equation reads
\begin{align}\label{LaxEq1Solution}\psi= \psi_0(k,t)e^{k x},
\end{align}
where the amplitude $\psi_0(k,t)$ is a function of wave number and time.

Substituting $\psi$ in the second Lax equation \eqref{Lax2},
we obtain
\begin{align}\partial_t\psi=\dfrac{\eta^2}{\tau}\partial_{xx}\psi +\dfrac{V(x)}{\tau\eta}\psi +\dfrac{e_0}{\tau}\phi\psi.
 \end{align}

This equation describes the time evolution of the scattering amplitude $\psi_0(k,t)$.
 Using the asymptotic $V(x)\rightarrow0$ at $x\rightarrow -\infty$, we obtain
\begin{align}\partial_t\psi_0(k,t)=\dfrac{\eta^2}{\tau}k^2 \psi_0(k,t) 
  \end{align}
and
\begin{align}\psi_0(k,t) =\psi_0(k,0)e^{\frac{k^2\eta^2}{\tau}t}.
 \end{align}

Hence the solution of the forward problem has the form
 \begin{align}\label{forward}
\psi(x,t)=\psi_0(k,0)e^{k x}e^{\nu t},
 \end{align}
where $\nu=\frac{k^2\eta^2}{\tau}$.

Using the solution \eqref{forward}, we  solve the inverse scattering problem. First, we define the function $F(x,t)$ as 
\begin{align}
F(x,t)=F_0e^{k x}e^{2\nu t}
\end{align}
and search the solution of eq.~\eqref{GLM}  in the form $K(x,y,t) = M(x,t) e^{k y}$.
After substitution we get
\begin{align}
&M(x,t) e^{k y}+F_0e^{k (x+y)}e^{2\nu t} \nonumber \\
&+\int\limits_x^{+\infty}M(x,t) e^{k z}F_0e^{k (y+z)+2\nu t}dz=0, \nonumber
\end{align}
then
\begin{align}
&M(x,t) +F_0e^{k x+2\nu t}  \nonumber \\
&+ M(x,t) F_0e^{2\nu t}\int\limits^x_{-\infty} e^{2k z}dz=0,\nonumber
\end{align}
and finally
\begin{align}
M(x,t) +F_0e^{k x+2\nu t} + M(x,t)F_0e^{2\nu t} \dfrac{e^{2k x}}{2k}=0.\nonumber
\end{align}

\begin{figure}[ht]
 \centering
\includegraphics[width=6.0cm]{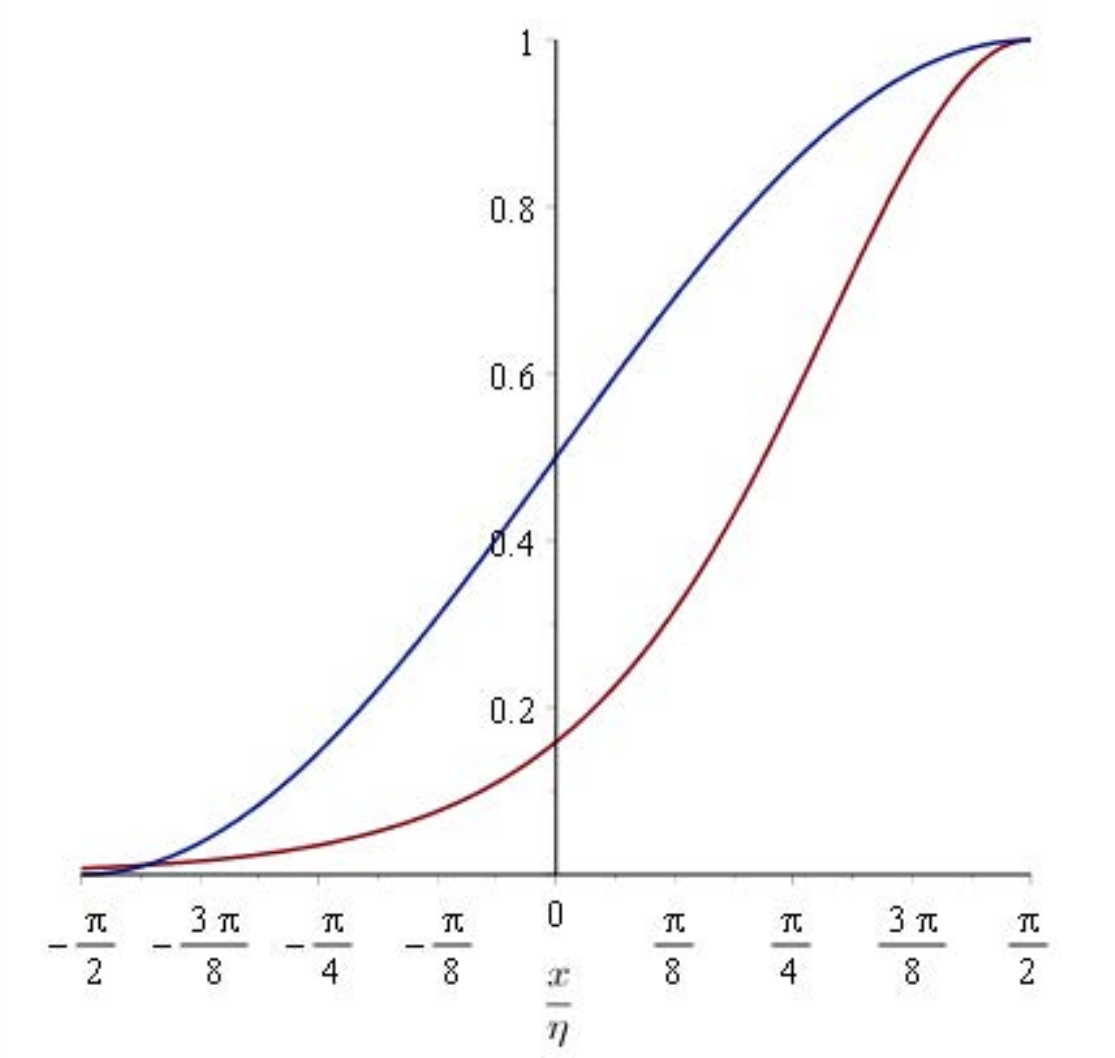}
\caption{Comparison of the functions $\frac{1}{2}\sin(\frac{ x}{\eta})+\frac{1}{2}$ (top, blue) and ${2k^2\rm {sech^2}}(-k x -\frac{\pi}{2})$ (bottom, red) for $k=\frac{1}{\sqrt{2}}$ and $\eta =\sqrt{2}$.} \label{SinSech}
\end{figure}

Hence
\begin{align}
M(x,t) =\dfrac{-2k F_0e^{k x+2\nu t}}{2k + F_0e^{2k x+2\nu t}}\nonumber
\end{align}
and from eq.~\eqref{Result}  we obtain
 \begin{align} \label{Result2}
 \phi(x,t) =-2k^2{\rm {sech^2}}(-k x - \nu t -\delta),
  \end{align}
  where $\delta =1/2 \log(F_0/2k)$.
This solution has the form of the known solution of KdV equation. The difference is in the dependency of $\nu$ on $k$, which is quadratic. The comparison with the solution \eqref{Sin_solution} gives $k=\dfrac{1}{\eta}$, $\nu=\dfrac{1}{\tau}$, $v=\dfrac{\eta }{\tau}$,  $ e = 1 $, $e_0=1$.
 The functions are compared in Figure \ref{SinSech} for $\eta = \sqrt{2}$ and $t=0$.

\subsubsection{Variant II}

The second variant of  the  Lax operators reads
\begin{eqnarray}\label{SecondVariant}
&\mathcal{L}&= \tau \eta\partial_x+ \tau \phi,\nonumber\\
&\mathcal{A} &= \dfrac{\eta e}{\tau}\partial_{x} +\dfrac{V(x)}{\tau\eta}-\dfrac{\eta}{\tau}\partial_x\phi.
\end{eqnarray}

The non-zero components of the Lax equation are
\begin{align}
&[\partial_x, V(x)]\psi= f_\phi(\phi)\psi;\nonumber\\
&[ \phi,\partial_{x}]\psi= \phi \psi_{x}-\phi_x\psi-\phi\psi_x=-\phi_x\psi;\nonumber\\
&[\partial_{x}, \phi_x]\psi=-\phi_{xx}\psi.
\end{align}

After substitution  in eq.~\eqref{LaxEq0}, we recover  the phase-field equation \eqref{pfe} with arbitrary $e$:
\begin{align}
&\tau \partial_{t}\phi =  \eta^2\partial_{xx}\phi- f_\phi(\phi) +\eta e\partial_{x}\phi.
\end{align}

The first Lax equation has the same form as eq.~\eqref{LaxEq1}
\begin{align}
\eta \partial_{x}\psi &=-  \phi\psi+k\eta\psi.
\end{align}
with the solution \eqref{LaxEq1Solution}. The second Lax equation reads
\begin{align}\partial_{t}\psi=\dfrac{\eta e}{\tau}\partial_{x}\psi +\dfrac{V(x)}{\tau\eta}\psi -\dfrac{\eta}{\tau}(\partial_{x}\phi)\psi.
 \end{align}

 Using the asymptotic $V(x)\rightarrow0$ and $\phi_x\rightarrow0$ at $x\rightarrow -\infty$, we obtain
\begin{align}\psi_0(k,t)_t=\dfrac{\eta e}{\tau}k \psi_0(k,t) ,
  \end{align}
which has the solution
\begin{align}\psi_0(k,t) =\psi_0(k,0)e^{\frac{k\eta e}{\tau}t} .
 \end{align}

Hence the solution of forward problem in the second variant is 
 \begin{align}\label{forward2}
\psi(x,t)=\psi_0(k,0)e^{k x}e^{\frac{k\eta e}{\tau}t}.
 \end{align}
  
 Here we obtain a linear dispersion relation, i.e. the frequency is proportional to $k$, $\nu=\frac{k\eta e}{\tau}$.

 The procedure of the solution of the inverse scattering
problem is similar to Variant I.
Finally, the solution is
 \begin{align} \label{Result_II}
 \phi(x,t) =-2k{\rm {sech^2}}(-k x - \nu t -\delta).
  \end{align}

Comparison with the sin-solution \eqref{Sin_solution} gives $k=\dfrac{1}{\eta}$, $\nu=\dfrac{e}{\tau}$ and $v=\dfrac{\nu}{k}=\dfrac{\eta   e}{\tau}$.

It can be shown that eq.~\eqref{Result_II} is the solution of PF equation \eqref{pfe} with the potential $\tilde f(\phi)=2\phi^2(1-\phi)$.  It is also interesting to mention here that with this potential the PF equation can be easily transformed to the KdV equation by taking the space derivatives of the two first terms on the right hand side of the PF equation. This kind of potential is not stable, because it is
unbounded from below and offers an infinite energy for a very large $\phi$. To overcome this difficulty, we suggest to use the absolute value $|1-\phi|$, which produces the second minimum and  allows to get a stable soliton. 

\begin{figure}[ht]
 \centering
\includegraphics[width=8.0cm]{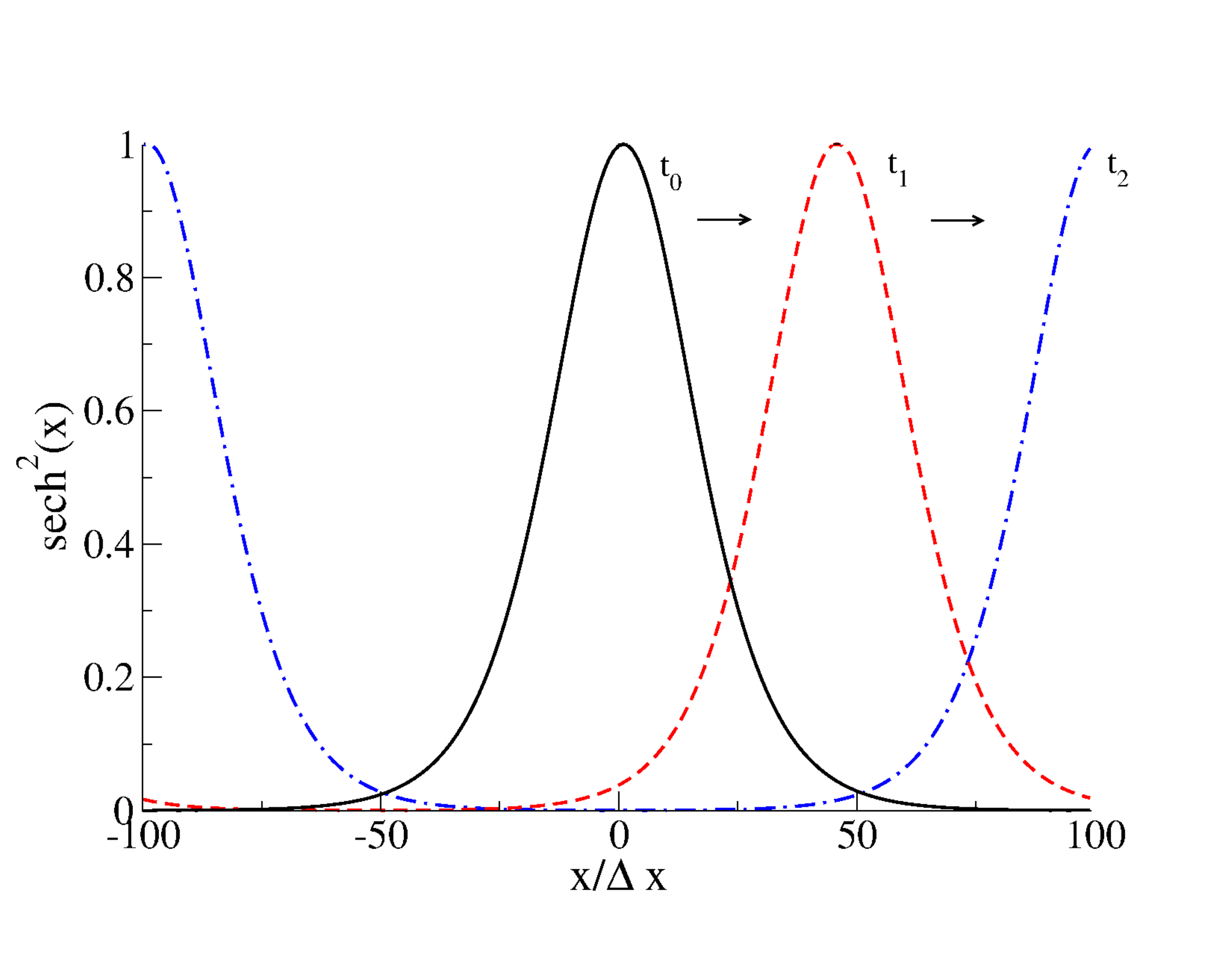}
\caption{Traveling wave solution for the functions ${\rm {sech^2}}\left(-x/\eta - t/\tau \right)$ with periodic boundary conditions. }\label{FigSech}
\end{figure}

Figure \ref{FigSech} shows the moving soliton solved numerically by the PF equation \eqref{pfe}
with $\tilde f(\phi) = 2\phi^2|1-\phi|$, the derivative of which is
\begin{eqnarray}\label{pfeSech}
\tilde f_\phi(\phi) = 2\phi\left[2|\phi-1|+\frac{\phi(\phi-1)}{|\phi-1|}\right].
\end{eqnarray}

The parameters of the model are: $\eta = 20$, $e=1$, $\tau = 0.5 $, $\Delta t =1$, $\Delta x = 1$.
The numerical solution in Figure~\ref{FigSech} sketches the analytical traveling wave solution 
\begin{align} \label{SechNumeric}
 \phi(x,t) ={\rm {sech^2}}\left(-\dfrac{x}{\eta} - \dfrac{t}{\tau} \right).
  \end{align}

 Note that the solution \eqref{SechNumeric} defines a symmetric soliton. Such kind of solitons are typically derived from a KdV type of equation which is 3rd order in the space derivative, while \eqref{SechNumeric} is derived from a 2nd order PF equation. The important difference in the type of soliton solutions, symmetric or antisymmetric, thereby seems to lie in the parity asymmetry of one operator, either the differential operator or the potential. Here more future work is necessary. The form \eqref{pfeSech} opens a new class of symmetric soliton solutions. 

\subsubsection{Variant III}

The third variant of  the  Lax operators is
\begin{eqnarray}\label{VariantIII}
&\mathcal{L}&= \tau \eta\partial_x+ \tau \phi,\nonumber\\
&\mathcal{A} &= \dfrac{\eta e}{\tau}\partial_{x} +\dfrac{1}{\tau\eta}V(x)-\dfrac{\eta}{\tau}\partial_x\phi +\dfrac{\gamma}{\tau},
\end{eqnarray}
where $\gamma$ is added to Variant II for the normalization of the scattering data, see \cite{Novikov1984}.  

Now we consider the imaginary forms of Lax equations. 
Substituting the operators and taking $\lambda=\tau\eta k$, we get the first  equation
\begin{align} \label{LaxEq11}
i\partial_x\psi + i\phi\psi/\eta=k\psi.
\end{align}
Assuming an asymptotic behavior $\phi\rightarrow0$ at $x\rightarrow\infty$, we obtain %
\begin{align}
i\partial_x\psi=k\psi
\end{align}
with the solution %
\begin{align}\label{LaxEq1Solution2}
\psi= e^{-ik x} + b(k,t)e^{ik x} \,\, \text{ for } x\rightarrow\infty,\nonumber\\
\psi=a(k,t)e^{-ik x}\,\,  \text{ for } x\rightarrow-\infty.
\end{align}

This solution  describes scattering from the right of the incident wave $e^{-ikx}$ on the potential $\phi$, $b(k,t)$ represents a reflection coefficient and $a(k,t )$ is a transmission coefficient.
The second Lax equation with the same asymptotic becomes
\begin{align}
i\partial_t\psi=\dfrac{\eta e}{\tau}\partial_{x}\psi+\gamma \psi.
 \end{align}

By substitution of $\gamma=ik\eta e$ and $a(k,0)=0$, we obtain 
\begin{align}b(k,t) =b(k,0)e^{\frac{i2k\eta e}{\tau}t}.
 \end{align}

Hence the solution of the forward problem  has the form 

 \begin{align}\label{forward4}
\psi(x,t)=b(k,0)e^{-ik x}e^{2i\nu t},
 \end{align}
where $\nu=\frac{k\eta e}{\tau}$. 

To solve the inverse scattering problem, we define  $F(x,t)$  as follows
\begin{align}
  F(x,t)=F_0e^{-i(k x-2\nu t)}.
 \end{align}

We search for a solution  in the form
\begin{align}
K(x,y,t) \sim M(x,t) e^{-ik y}. 
 \end{align}

After substituting it in eq.~\eqref {GLM},  results in 
\begin{align}
&M(x,t) e^{-ik y}+iF_0e^{-ik (x+y)}e^{2i\nu t}\nonumber\\
&+i\int\limits_x^{+\infty}M(x,t) e^{-ik z}F_0e^{-i(k (y+z)-2\nu t)}dz=0,\nonumber\\
&M(x,t) +F_0e^{-i(k x-2\nu t)} \nonumber\\
&+ M(x,t) F_0e^{i\nu t}\int\limits_x^{+\infty} e^{-2ik z}dz=0.
\end{align}

The last term cancels by adding the complex conjugate (c. c.) function. Hence
\begin{align}
M(x,t) =-F_0e^{-i(k x-2\nu t)},\nonumber\\
K(x,x,t)=-F_0e^{-i(2k x-2\nu t)}.\nonumber
\end{align}

As a result, we obtain from \eqref{Result}
 \begin{align} \label{Result20}
 \phi(x,t) &=- \frac{2F_0k}{i}e^{-i(2kx-2\nu t)} =-\frac{F_0}{i\eta}e^{-i\frac{(x-v t)}{\eta}},
  \end{align}
where 
\begin{align} \label{Lambda}
k=\frac{1}{2\eta}, \,\,\, \nu = \dfrac{e}{2\tau}, \,\,\, v=\dfrac{\nu}{k}=\dfrac{\eta e}{\tau}.
  \end{align}

  By adding the c.~c. function and normalizing, a real solution can be found in the form \eqref{Sin_solution}.

\subsection{Lax pair for the relativistic case}\label{Relativ}

The phase-field equation for the relativistic singularity can be written as 
\begin{eqnarray}\label{pfeR}
\tau \partial_{t}\phi = \eta^2\left(\partial_{xx}\phi - \frac{1}{c^2}\partial_{tt}\phi\right)+ f_\phi(\phi)+\eta e\partial_{x}\phi. 
\end{eqnarray}

This equation has the form of the Klein-Gordon equation with advection. Using the solution that the singularity moves with the velocity $v= \eta e/\tau$ according to advection equation, we can replace $\phi_{tt}$ by   $v^2\phi_{xx}$ \cite{Steinbach2017zn}. Then the equation becomes
\begin{eqnarray}\label{pfeR2}
\tau \partial_{t}\phi = \eta_v^2\partial_{xx}\phi + f_\phi(\phi)+\eta e\partial_{x}\phi,
\end{eqnarray}
where  $\eta_v = \eta \sqrt{1-v^2/c^2}$.

The Lax pair can be chosen based on the second variant \eqref{SecondVariant} 
\begin{align}\label{SecondVariantKG}
&\mathcal{L}= \tau \eta_v\partial_x + \tau \phi,\nonumber\\
&\mathcal{A} = \dfrac{\eta e}{\tau}\partial_{x} +\dfrac{1}{\tau\eta_v}V(x)-\dfrac{\eta_v}{\tau}\phi_x.
\end{align}

With this choice the equations for the pilot-wave function become
\begin{align}
&\eta_v\partial_{x}\psi_x= -\phi\psi  + \psi,\\
&\tau\partial_{t}\psi=\eta e\psi_x  +\frac{V(x)}{\eta_v}\psi-\eta_v\partial_{x}\phi  \psi.
\end{align}

\section{Quantum phase field concept }\label{section_QPF}

Let $\phi$ be a scalar quantum field which acts on a quantum state $\psi =a e ^{-i\theta}$ of a vacuum oscillation. The Hamiltonian of such a system, which formally corresponds to a free energy of a closed thermodynamic system, can be expanded in the non-relativistic case (for the general case see \cite{Steinbach2017zn})
\begin{align}
H &=U_0\int_\Omega \langle\psi^\dagger| \hat h |\psi\rangle \,dx, \label{qpfe}\\
\hat h(\phi)&=\frac{\eta^2}{2}(\partial_x\phi)^2+ f(\phi) \nonumber
\end{align}
as the integral over the domain $\Omega$ of the free energy density, which is defined as the expectation value of the non-linear soliton operator $\hat h(\phi)$ acting on the pilot wave $\psi$.
The first term can be calculated as
\begin{align}
   \langle \psi^\dagger | \partial_x \phi \partial_x\phi |\psi\rangle &= \partial_x(\phi \psi^\dagger)   \partial_x(\phi \psi) \nonumber\\
 &=\left(\psi^\dagger \partial_x\phi+\phi \partial_x\psi^\dagger\right)\left(\psi \partial_x\phi+\phi \partial_x\psi\right)
 \nonumber\\
 &=\psi\psi^\dagger \left(\partial_x\phi\right)^2+\phi^2\partial_x\psi\partial_x\psi^\dagger \nonumber\\
 &+ \phi  \partial_x\phi\left(\psi\partial_x\psi^\dagger+\psi^\dagger\partial_x\psi \right).
\end{align}

The integration of the last term gives 0. Then by the relaxation dynamics
\begin{align}\label{Relax}
 \psi\psi^\dagger\tau\partial_{t} \phi  =-\frac{1}{U_0}\frac{\delta H}{\delta \phi} ,
\end{align}
we obtain
\begin{align}\label{QPF1}
\psi\psi^\dagger\tau\partial_{t} \phi =\eta^2\psi\psi^\dagger\partial_{xx}\phi -\eta^2\phi\partial_{x}\psi^\dagger\partial_{x}\psi +\psi\psi^\dagger f_\phi(\phi).
\end{align}

Substituting $\partial_{x}\psi^\dagger\partial_{x}\psi=(\theta_x)^2\psi\psi^\dagger$ and $f_\phi=\phi-\dfrac{1}{2}$ as defined in \eqref{dop2}, we get the PF equation with the DO potential 
\begin{align}\label{QPF2}
 \tau\partial_{t} \phi &=\eta^2\partial_{xx}\phi + f_\phi- \eta^2 (\theta_x)^2\phi,
\end{align}
where $\eta^2(\theta_x)^2$ is a quantum driving force. Using the complex function $\phi=e^{-i\theta(x)}$ with $\theta_x =1/\eta$, eq.~\eqref{QPF2} can be rewritten as
\begin{align}\label{QPF3}
 i\tau\partial_{t} \phi &=\eta^2\partial_{xx}\phi +   f_\phi + i \eta\partial_x\phi.
\end{align}

Equation~\eqref{QPF1} has the similar meaning as eq.~\eqref{PFeqLax} which is obtained by means of the Lax pair method. The function  $\psi\psi^\dagger$ is an analogy of the function $\psi$ in section \ref{SectionLax}.  Hence, using the Lax method, we can find the Lax equations for $\psi$ and $\psi^\dagger$. For eq.~\eqref{QPF3}  we can choose the following Lax pair
\begin{align}\label{LaxQPF1}
\mathcal{L} &= \tau \eta\partial_x+ \tau \phi,\nonumber\\
\mathcal{A} &= \dfrac{\eta^2}{\tau}\partial_{xx} +\dfrac{V(x)}{\tau\eta}+\dfrac{i}{\tau}\phi,
\end{align}
where $V(x) = -\int^{x} {f_\phi(\phi)\,dx}$. 
The non-zero components of the Lax equation are
\begin{align}
&[\partial_x, V(x)]\psi= -f_\phi(\phi)\psi;\nonumber\\
&[ \phi,\partial_{xx}]\psi=-\phi_{xx}\psi-2\phi_x\partial_x\psi;\nonumber\\
&[ \partial_x,\phi]\psi=\phi_{x}\psi.
\end{align}
This is valid also for $\psi^\dagger$.
Then we can reproduce the PF equation \eqref{QPF3} for the field $\phi$ by the Lax equation $i\mathcal{L}_t+[\mathcal{L},\mathcal{A}]=0$. 

The corresponding Lax equation for the complex function $\psi$ (or $\psi^\dagger$) have the form $i\psi_t =\mathcal{A}\psi$, i. e.
\begin{align}\label{LaxQPF2}
i\tau \partial_{t}\psi= \eta^2\partial_{xx}\psi +\dfrac{V(x)}{\eta}\psi+ i \phi\psi.
\end{align}

An alternative method to obtain the evolution equation for a state function is the relaxation dynamics:
\begin{align}\label{Relax2}
 i\tau\partial_{t} \psi  &=-\frac{1}{U_0}\frac{\delta  H}{\delta \psi^\dagger} \nonumber\\
 &=\frac{\eta^2}{2}\phi^2\partial_{xx}\psi-\frac{\eta^2}{2}(\partial_{x}\phi)^2\psi- f(\phi)\psi.
\end{align}

From this equation we can see that the $\psi$ function exists only in  the  region where $\phi>0$ and it is bounded by the potential $f(\phi)$, which tends to 0 at $\phi\rightarrow0,1$. Therefore, the problem is similar to the Schr\"odinger equation for a particle in a box potential (see below, Figure \ref{doublon_welle} ).

\section{The doublon in 1+1 dimensions}\label{doublon11}

Having the solution for the wave  $\phi$ in the form of the so-called half sided `soliton' with spinor character regarding parity in space, we easily construct a double sided `soliton', which we will call `doublon' as an antisymmetric pair of two half sided solitons and the `space' in between them: 
\begin{align}
\label{travellingWave}
\phi =\begin{cases} 
\dfrac 12\,\,+ \,\, \dfrac 12&\sin\left(\dfrac{\pi(x-x_1+vt)}{\eta}\right)\\
&\mbox{for}\;\; -\frac{\eta}2\le x-x_1 +vt< \frac{\eta}2 \\
\dfrac 12\,\,- \,\, \dfrac 12& \sin\left(\dfrac{\pi(x-x_2-vt)}{\eta}\right)\\
&\mbox{for}\;\; -\frac{\eta}2\le x-x_2 -vt< \frac{\eta}2 \\
 1 & \mbox{for}\;\; x_1-vt+\frac{\eta}2\le x < x_2+vt-\frac{\eta}2 \\
0 & \mbox{otherwise.}
\end{cases}
\end{align}


\begin{figure}[ht]
 \centering
\includegraphics[width=8.3cm]{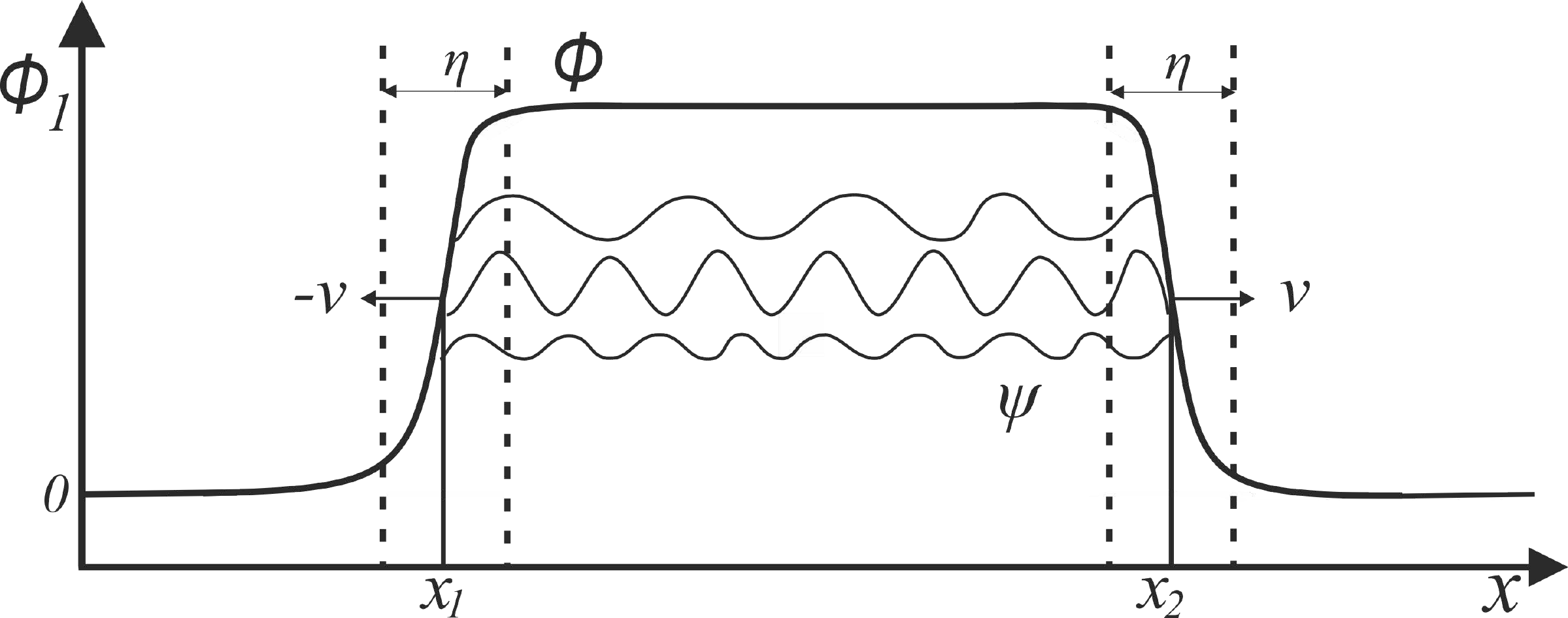}
\caption{The doublon $\phi$ as a combination of a left- and a right-moving soliton defines the box for vacuum fluctuations $\psi$ which determine the spacial energy of the doublon. }\label{doublon_welle}
\end{figure}
 
The solution is depicted in Figure \ref{doublon_welle} in 1 dimension neglecting time. 
The doublon consists of an antisymmetric pair of two half-sided solitons and a finite space, attributed by to a 1-dimensional line coordinate $x$, where $\phi(x) \equiv 1$. This space is bounded by the non-local transition regions of width $\eta$ where $\frac \partial {\partial x} \phi(x) \ne 0$, the right- and left moving soliton, which we identify with particles moving with relative velocity $v=v_{\rm p}$. Thereby the doublon forms a 1-dimensional box for vacuum fluctuations $\psi$. Regions outside the doublon, i.e. regions on the line coordinate $x$ with $\phi(x) \equiv 0$ have no physical meaning. 
The doublon forms the elementary building block of the physical world in the present concept. It unifies space-time and mass in a monistic structure. `Mass', attributed by positive energy, is related to $\eta^2$ while space is attributed by negative energy $e$. The latter is easily calculated from quantum fluctuations with discrete spectrum $p$ and frequency $\omega_p= \frac {\pi c p }{2\Omega}$, where $\Omega=|x_1-x_2|$ is the size of the doublon. According to Casimir \cite{Casimir1948}, this has to be compared to a continuous spectrum. This yields the negative energy $e$ of space:
\be\label{EC}
e = \alpha \frac {h c}{4\Omega} \left[ \sum_{p=1}^\infty p - \int_1^{\infty} p dp \right]\\ = - \alpha \frac {h c}{48\Omega},
\ee
where $\alpha$ is a positive, dimensionless coupling coefficient. We have used Euler--MacLaurin formula in the limit $\epsilon \rightarrow 0$ after renormalization $p \rightarrow p e^{-\epsilon p}$. The quantum fluctuations are the Schroedinger type solutions of the $\psi$ wave in the dBB picture. Here we have to note one important difference in the approaches (see also the discussions in chapter \ref{discussion}). In the dBB program the pilot wave is the primary object which lives in a given environment and drives the $u$ wave in the form of a soliton as the object of investigation. In the present concept, the  $\phi$ wave in the form of the doublon defines the environment for the $\psi$ wave. Up to now only the `quasi-static' limit has been worked out, i.e. that the doublon solution is kept fixed for the quantum solution \eqref{EC}. The direct coupling of both waves by their phase in the transition region $0<\phi<1$ may be investigated using the Lax formalism. We leave a closer investigation of the coupling between the  $\phi$ and $\psi$ waves and a dynamic coupled solution to future work.


%

\section{The doublon network in 1+1+2 dimensions}\label{section_doublon}

To proceed towards a multi-dimensional (in the space coordinate) case we have to 

\begin{itemize}
\item define a set of doublons $\phi^I, I=1...N$ with a number $N > 6$ for a 3-dimensional space filling network,
\item define the interaction between doublons and 
\item embed the description in 3+1 dimensional space time. 
\end{itemize}

The second item will follow canonically from the postulate that the set of doublons is closed in itself, i.e.

\be\label{closed_set}
\sum_{I=1...N} \phi^I = 1.
\ee

Using expression (\ref{qpfe}) for the free energy of an individual field $I$,  $  H\rightarrow  H^I$, we define the total  free energy $\ H=\sum_{I=1...N}  H^I$, then conserving (\ref{closed_set}) (see \cite{Steinbach1999} for details) we end up with the equation of motion for the doublons

\bea \label{mpf}
\nonumber \tau \partial_t \phi^I
&=& - \frac 1 {N } \sum_{J=1...N} \{\frac \delta{\delta \phi^I} - \frac \delta{\delta \phi^J} \} \frac { H}{U_0} \\
&=& - \frac 1 {N} \sum_{J=1...N} \Phi^{IJ}.
\eea

The last expression defines an antisymmetric object of dual character, the pair-exchange operator $\Phi^{IJ}$ between two doublon fields $I$ and $J$. This beautiful result is a mere consequence of the system being closed in itself and leads to a natural decomposition of the multi-body interaction between the doublons into pair-wise contributions. Knowing the structure of doublons  consisting of two antisymmetric solitons and the space between them, it is obvious that the interaction of doublons only happens in the soliton regions where $0<\phi^I<1$ with size $\eta$. This size is estimated in \cite{Steinbach2017zn} to be below $10^{-15}$ m, i.e. in the range of elementary particles in the classical sense. We will call this region `quasi-local', meaning that we have a non-local theory with highly localized states. Within the quasi-local position of an elementary particle all $N$ doublon fields may interact, the particle is understood as a junction between doublons. Although doublon fields are constructed along a 1-dimensional line coordinate, we may argue that the junctions form 0-dimensional object.  Later, when introducing charges in section \ref{charge}, we will also discuss a finite width of the doublons to allow for optical excitations. Each pair-exchange operator $\Phi^{IJ}$ carries an orientation from the half sided soliton at the endpoint of the doublon $\phi^I$, considered as a spinor. Due to the isomorphism of spinors with the 3-dimensional SU(2) group, we therefore may embed the doublons into a 3-dimensional spacial environment within the quasi-local environment of a particle and a small surrounding.  Here we define the 3+1 dimensional field variable $\tilde\phi^I(\vec x, t)$ for which a standard phase-field description in 3-dimensional space is applicable. Outside this local environment the field collapses to the space like part of a doublon. We formally define the doublon as the trajectory, or path, of a classical flow between the endpoints of the doublon in space time.  The advancing flow corresponding to doublon $I$, $\Xi_{Adv}^I$ and its time reversed equivalent, the retarded flow  $\Xi_{Ret}^I$, connects space-time events $(\vec x_1,t_1)$ and $(\vec x_2,t_2)$. This path is determined from a minimum action condition $E(\tilde \phi^I) = 0$ (see \cite{Mazenko2011,Bartelmann2016}):

\be
 \tilde \phi^I(\vec x, t) = \Xi_{Adv}^I  \tilde \phi^I(\vec x_1, t_1); \;\;\;  \tilde \phi^I(t,\vec x) = \Xi_{Ret}^I  \tilde \phi^I(\vec x_2,t_2)
\ee

The doublon then follows the probability $P$ to find non-zero values of the field $ \tilde \phi^I(\vec x, t)$ at space point $\vec x$ and time $t$ in 3+1 dimensions :

\be\label{prob}
 P\left[ \tilde \phi^I(\vec x, t), \tilde \phi^I(\vec x_1, t_1)\right]  = \mathcal{\int}\mathcal{D}  \tilde \phi^I \; \delta_D   \left[E(\tilde \phi^I)\right],
\ee
where $\delta_D$ is the Dirac functional delta function. 

 Any event $(\vec x,t)$ laying of the doublon connecting the events $(\vec x_1,t_1)$ and $(\vec x_2,t_2)$ can be reached either via the action of advancing operator on $(\vec x_1,t_1)$ or equivalently the retarding operator on $(\vec x_2,t_2)$.

\begin{figure}[ht]
centering
\includegraphics[width=8.0cm]{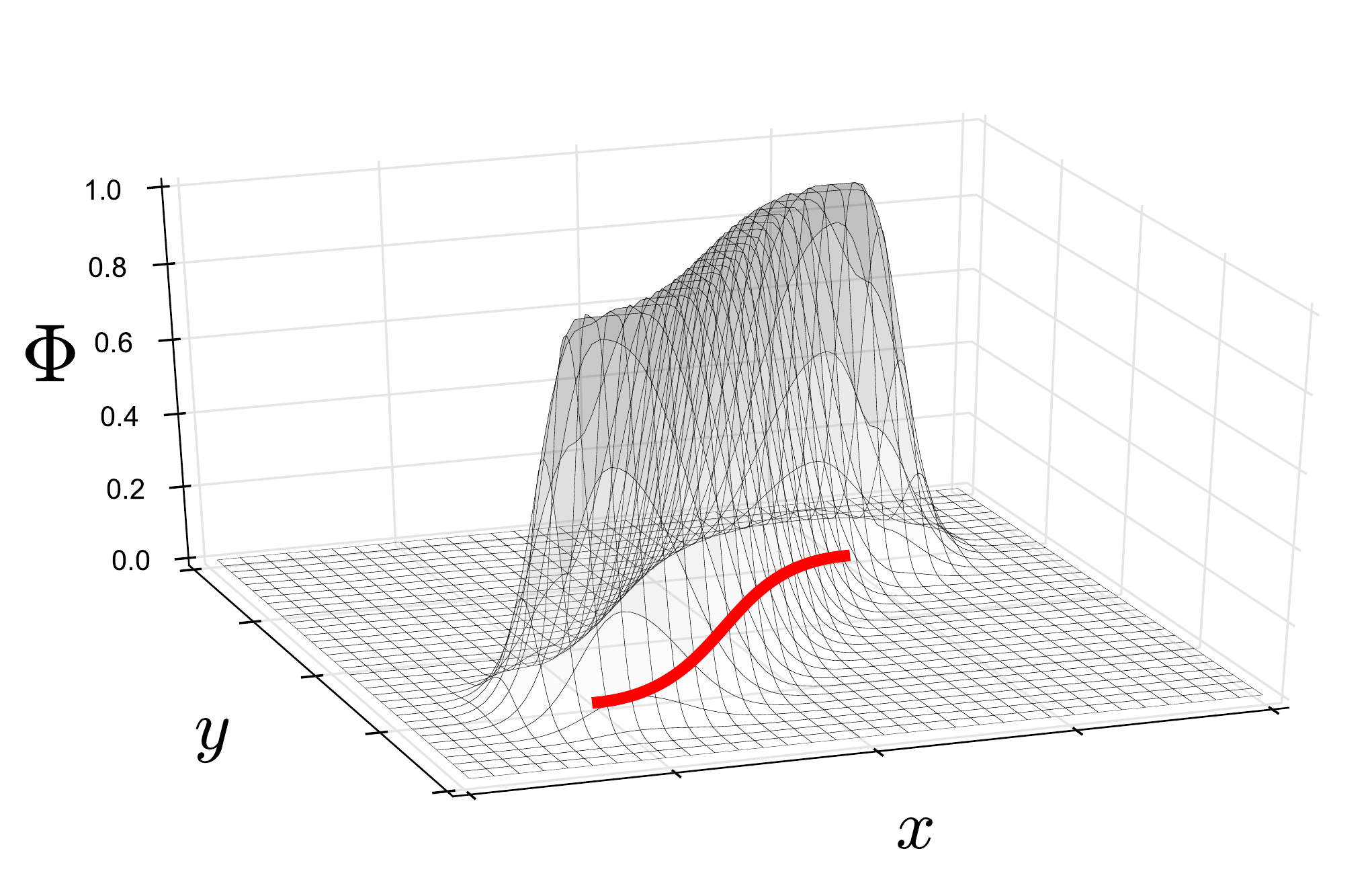}

\caption{Sketch of a doublon represented by the 3+1-dimensional field $\tilde\phi^I(\vec x, t)$ }\label{3D_doublon}
\end{figure}

\begin{figure}[ht]
\centering
\includegraphics[width=5.5cm]{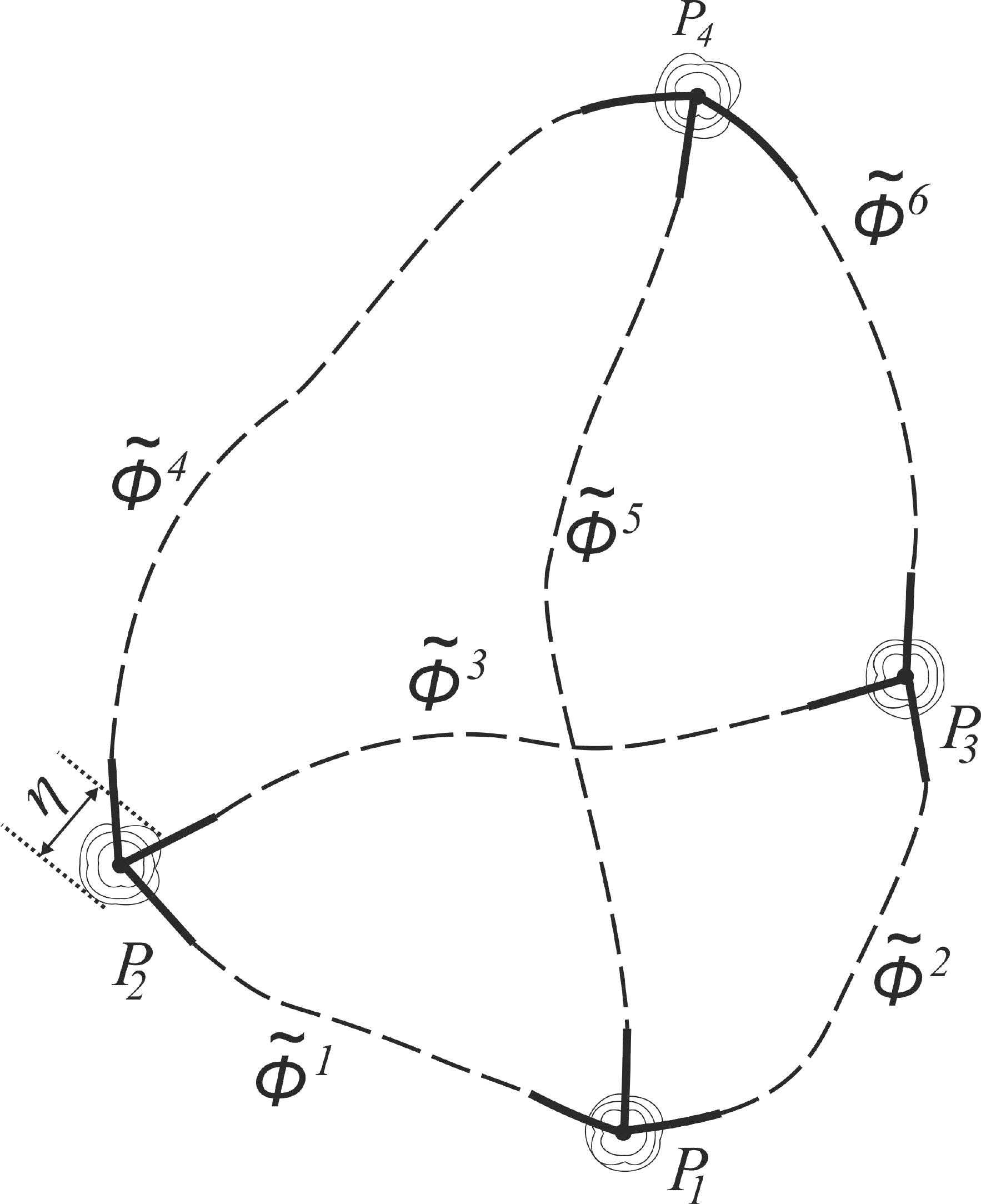}
\caption {Doublon network for $6$ interacting doublons $\tilde \phi^1 ... \tilde \phi^6$ forming a 3-dimensional space filling environment with $4$ particles $P_1 ...P_4$. The junctions of size $\eta$ are stretched out by a minimum of $3$ doublons $\phi^I$ defined on linear independent directions in 3-dimensional space.  }
\label{network}
\end{figure}

The doublon network is schematically depicted in Figure \ref{network} for $6$ interacting doublons forming a 3-dimensional space filling environment with $4$ particles.  As indicated,  we define around the particles a 3-dimensional environment of size $\eta$, corresponding to the (inverse) gradient of the fields $\phi^I$ which interact in this quasi-local junction. Since each doublon $\phi^I$ carries a different orientation in the 3-dimensional space it is embedded in, we define here the 3-dimensional field $\tilde \phi^I(\vec x, t)$. 
As indicated by the solid end-segments of the doublons, the volume of the junctions can uniquely be divided into volume segments assigned to the volumetric fields $\tilde \phi^I$. Outside the junctions these volume fields collapse to tube-like objects along the definition \eqref{prob}, as indicated by the dashed lines representing the space part of the doublon.

\section{Phase-field equation in electromagnetic field}\label{charge}

We will proceed here in the canonical way along the Ginzburg-Landau free energy functional for a superfluid phase defined by a doublon $\tilde \phi^I$ in electromagnetic field. The energy functional \eqref{qpfe} now reads:%
\begin{align}\label{qpfeA}
 H^I &=U_0\int_\Omega h^I( \tilde\phi^I, \mathbf{A}) \,d\mathbf{x},
\end{align}
where
\begin{align}
 & h^I( \phi^I, \mathbf{A}) = \frac{1 }{2} \left[\eta^2 n_s\left|\left(  \nabla
- \frac{q \mathbf{A}}{\hbar  }\right)\tilde \phi^I \right|^2  \right. \nonumber\\
   &\left.+ \, n_s| \tilde\phi^I(1-  \tilde\phi^I)| +\frac{1}{4}\frac{(\nabla n_s)^2}{ n_s}(\tilde \phi^I)^2+\frac{B^2}{\mu U_0   } \right] .
\end{align}

Here $\mathbf{A}$ is the vector potential, $B$ is the magnetic field, $\mu$ is the magnetic permeability, $\mathbf{x}$ is the space vector, $q$ is the charge, $U_0=  mc^2$ is the energy of  a superfluid, $n_s=\psi\psi^\dagger$ is the concentration of a superfluid,  $\eta = \dfrac{ \hbar }{ mc} $  is the interface width,  and $\dfrac{B^2}{2\mu}$ is the  energy  density of the magnetic field. 

Using the relaxation dynamic \eqref{Relax} with $\tau = \dfrac{ \hbar n_s}{ mc^2}$, we obtain
 the evolution equation for a phase in an electromagnetic field 
\bea
\nonumber  \tau \partial_t \tilde\phi^I&=\eta^2\left(  \nabla   - \dfrac{q \mathbf{A}}{\hbar  }\right)^2 \tilde\phi^I
   +( \tilde\phi^I- \dfrac{1}{2}) \\
&+\dfrac{1}{4}\dfrac{( \nabla n_s)^2}{ n_s^2} \tilde\phi^I.
\eea

In order to derive the kinetic equation for the superfluid, we minimize the functional $ H^I $ with respect to $\mathbf{A}$. First, we rewrite functional to collect all terms depending on  $\mathbf{A}$ as
\begin{align}
 & h^I(\tilde \phi^I, \mathbf{A}) =     h^I_0 (\tilde\phi^I) - \frac{\hbar q  n_s}{m  } ( \tilde\phi^I)^2\nabla \cdot \mathbf{A} \nonumber  \\ 
 &-\frac{\hbar q  n_s}{m  }\mathbf{A} \cdot\nabla  (\tilde\phi^I)^2+ \frac{ q^2  n_sA^2}{2m  }   (\tilde\phi^I)^2+ \frac{B^2}{2\mu}   .
\end{align}

 Since $\nabla \cdot \mathbf{A}=0$, the term with the divergence of $\mathbf{A}$ can be neglected.
By substitution $\nabla\times \mathbf{A} = \mathbf{B}  $, the variation of the last term with
respect to $\mathbf{A}$ gives:
\begin{align}
\delta {B^2} &=  \delta |\nabla   \times \mathbf{A}|^2= \frac{\partial |\nabla   \times \mathbf{A}|^2}{\partial (\nabla   \times \mathbf{A}) } \delta(\nabla   \times \mathbf{A})\nonumber\\
&=2(\nabla   \times \mathbf{A})\delta (\nabla   \times \mathbf{A}) \\
&= -2\nabla((\nabla   \times \mathbf{A})\times  \delta\mathbf{A})+ 2(\nabla \times \nabla\times \mathbf{A})\delta \mathbf{A},\nonumber
\end{align}
where  the integral of the first term is 0 by the Gauss' theorem.

Finally, the minimization of $  H^I $  gives
\begin{align}
 \frac{\delta   H^I}{\delta \mathbf{A}} =   - \frac{\hbar q n_s }{ m  }  \tilde\phi^I\nabla \tilde \phi^I + \frac{ q^2 n_s\mathbf{A}}{m   }  \tilde\phi_I^2+ \frac{\nabla\times  \mathbf{B} }{\mu  }  =0.
\end{align}

Substituting $   \nabla\times  \mathbf{B}  = \mu\,\mathbf{j}_s$,
we obtain the second London equation
\begin{align}
\mathbf{j}_s = -\dfrac{ q^2 n_s}{m} \mathbf{A}
 \end{align}
for the case of the uniform $ \tilde\phi^I=1$.

\section{Discussion and conclusion}\label{discussion}

\begin{figure}[ht]
\centering
\subfloat[-------------------------------------------------------]
{\includegraphics[width=6.0cm]{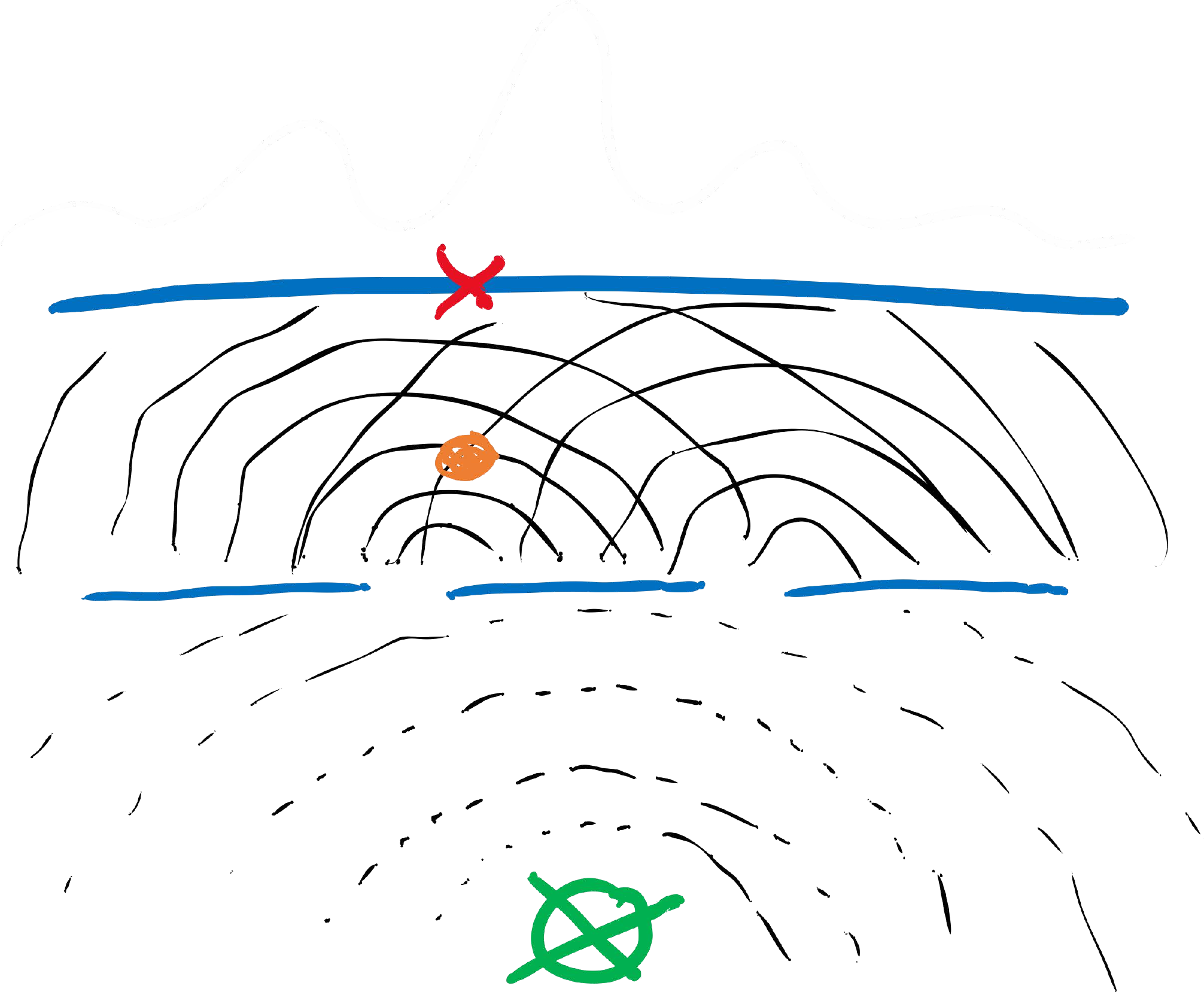}}
\vspace{0.1cm}
\subfloat[-------------------------------------------------------]
{\includegraphics[width=6.0cm]{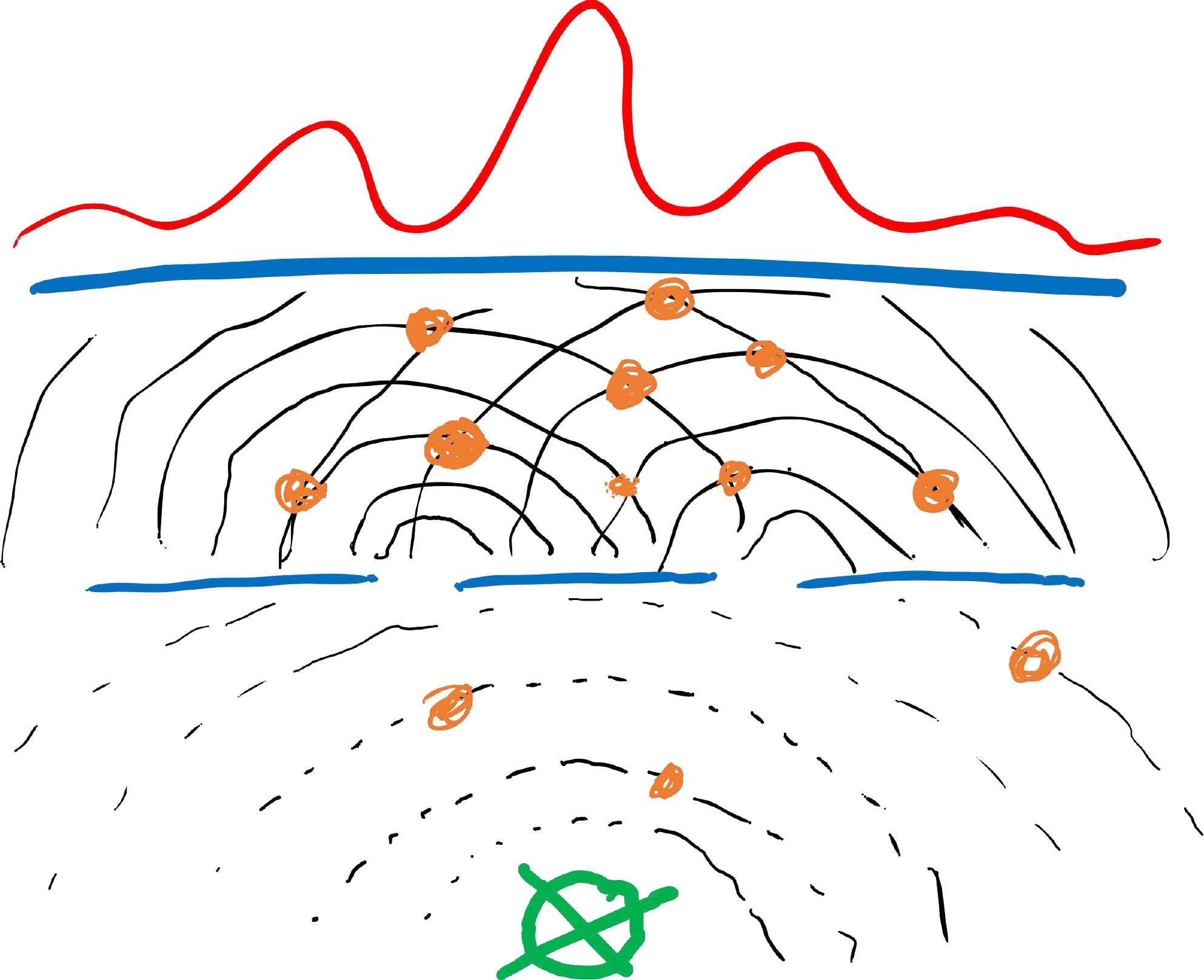}}
\vspace{0.1cm}
\subfloat[-------------------------------------------------------]
{\includegraphics[width=6.0cm]{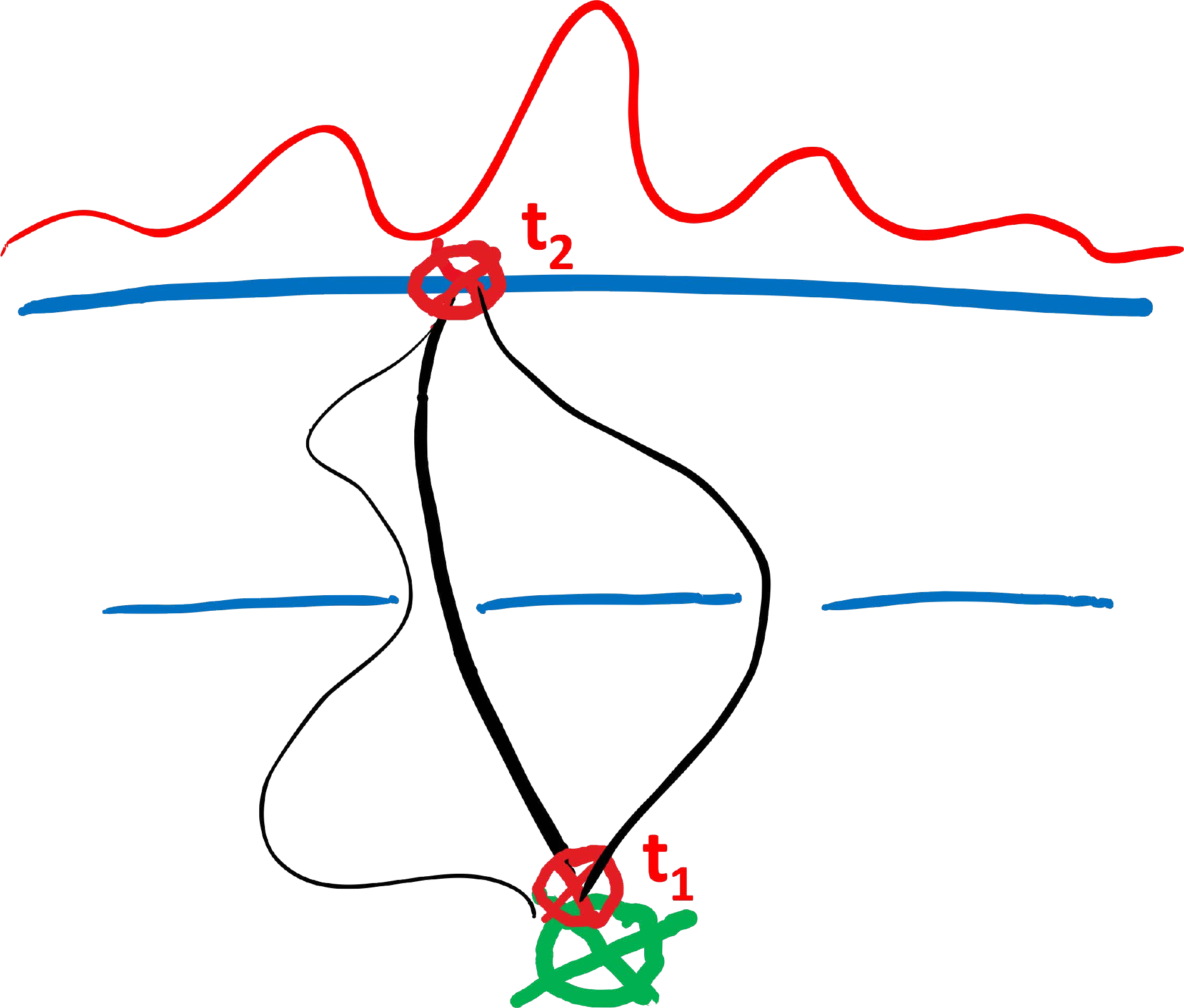}}
\caption{Double slit experiment. a) probability wave with one individual particle, b) probability wave with many particles, c) path integral for a single particle trajectory}\label{double_slit}
\end{figure}

The de~Broglie-Bohm double solution program sets a framework of coupled wave equations to represent `particles' which are guided by a probability wave function in order to be consistent with observations of quantum statistical behavior of these particles. Let us recall the famous double slit experiment: particles (photons or electrons or even fullerenes \cite{Gerlich2011}) emitted on one side of a double slit  form an interference pattern at the absorber screen. It must be stated clearly that an individual particle has only one strike on hte absorber screen. The interference pattern, however, also appears if the objects are recorded with some time delay, i.e. that the events can be treated uncorrelated. The situation is depicted in Figure \ref{double_slit} a) and b). An individual particle only has one strike at the absorber screen, while a number of particles form the interference pattern. In the dBB picture the pilot wave replaces the probability wave in the Copenhagen interpretation. From a rational point of view these waves can be seen as kind of a Fourier transformed snapshot of the experimental set up. Time needed for transport is not included in this interpretation. Also the scale on which interference is observed only depends on the relation between the distance of the slits and the characteristic wave length of the pilot wave, i.e. the dominating Fourier component. This solution is a classical wave solution negating the existence of discrete objects, the particles. It can be obtained by solving the differential wave-mechanical equation for given boundary conditions, which represent the environment: emitter, absorber screen and double slit barrier. So it is not astonishing that the slit experiment can be repeated with water waves and small cork particles in the classroom, or in a more sophisticates set up with oil droplets in the laboratory \cite{Couder2005,Couder2006}. Compare also the hydrodynamic interpretation of quantum-mechanics by Madelung in the 1920th \cite{Madelung1926}.

Now we must recall one major criticism of Einstein (and others) on these interpretations of quantum mechanics that they anticipate an instantaneous knowledge of the situation in the experiment. The time needed for transport is not considered. An elegant solution to this problem is given by Feynman's path integral theory \cite{Feynman1948}, Figure \ref{double_slit} c). The probability to detect a particle at a special position at the absorber screen is determined by the action of the particle along different paths. A global wave within the whole experimental setup does not exist and the time of transport is consistently considered.  Also the scale, on which quantum interference can be observed, is determined by the action in comparison to Plancks quantum. So one can clearly distinguish between   quantum effects and classical hydrodynamic effects \cite{Couder2005,Couder2006}. 

The path integral picture leads us directly to the doublon-network model in section \ref{section_doublon}, where the doublon represents a kind of elementary space quantum connecting emitter and absorber. The doublon connects these points by a flow of quantum fluctuations. Elementary particles by themselves are defined from gradient contributions, half-sided solitons, on both ends of the doublon. The concept is consistent with the approach of dBB, but distinct in several aspects, the most important is the topology of space and the definition of particles.  

As we can see from section II, the PF formalism allows to treat both forms of the particle representation: the symmetric solution and the half-sided solution (Figure \ref{soliton}). Due to their parity one may identify the symmetric solution with a bosonic particle and the antisymmetric, or half-sided solution with a fermionic particle. We state that fermionic particles always have to come in pairs and that there is a conservation constraint. For bosonic particles, on the other hand, no conservation is expected since their creation or annihilation does not change the global wave solution apart from the individual position of the bosons. We may speculate that the space-doublons $\tilde \psi^I$ possess a transverse width which may be determined by a bosonic wave solution in transverse direction according to Eq. \eqref{pfeSech} and \eqref{SechNumeric}. In this transverse direction optical excitations are possible corresponding to electromagnetic waves. More future work is needed to investigate such solutions.

In section II, we have considered several variants of PF equations within the framework of the dBB double solution program. Special  equations are found which define particle-like wave solutions. More technically, we have separated the real parts of the PF equation, which are responsible for the phase evolution, and the imaginary parts for the amplitude evolution.  
A promising variant seems to be the definition of the particle velocity  as the velocity of an amplitude of a superwave whose phase is the probability wave function (see section \ref{Section_DeBroglieEquations_Pilot}).  Both, the amplitude and the phase, are solutions of Schr\"odinger-type equations, which have the form of the transformed PF equation. Hence, the particle is defined as a wave function which is coupled to the probability wave function through the superwave. If the state function changes due to a quantum or other forces, then the particle will change its velocity which depends on the phase of the probability function, whereas the quantum force is defined by the amplitude of the probability function. Therefore, the probability function works as the pilot wave in analogy the de~Broglie-Bohm theory.

Finally we have demonstrated the potential of the Lax formalism with the given explicit forms of the Lax operators to trace the coupling of the the particle wave $\tilde \psi$ and the probability wave $\phi$ in section II, as well as the coupling of the the quantum oscillations $\tilde \psi$ and the doublon $\phi^I$ in section IV.

In conclusion, our findings provide a new interpretation of particles in quantum mechanics and new opportunities for the formulation of the quantum wave equations.  Our treatment goes significantly beyond the original dBB program by establishing a consistent set of wave equations derived by variational principles. Furtheron it allows predictions (see \cite{Steinbach2017zn}) regarding structure formation in the universe and its accelerating expansion \cite{Riess1998}.

 \section*{Acknowledgement} The author would like to thank Fathollah Varnik, Bochum,  for support with discussions and suggestions and Dmitry Medvedev, Bochum/Novosibirsk, for critical reading of the manuscript.


\bibliographystyle{unsrt}

\end{document}